\documentclass[a4paper,usenatbib]{mnras}

\usepackage[T1]{fontenc}
\usepackage{ae,aecompl}

\usepackage[dvipsnames]{xcolor}

\usepackage{graphicx}	
\usepackage{amsmath}	
\usepackage{amssymb}	
\usepackage{BibDef}
\usepackage{ragged2e}
\usepackage{booktabs}
\usepackage{multirow}
\usepackage{mathrsfs}
\usepackage{comment}
\usepackage{ulem}
\newcommand{\msun}{\ensuremath{\, \mathrm M{\sun{}}}}

\title[Massive black hole pairing in the young Universe]{Global torques and stochasticity as the drivers of massive black hole pairing in the young Universe}

\author[E. Bortolas et al.]{%
Elisa Bortolas,$^{1}$\thanks{E-mail: elisa.bortolas@uzh.ch}
Pedro R. Capelo,$^{1}$
Tommaso Zana,$^{2}$
Lucio Mayer,$^{1}$ \newauthor
Matteo Bonetti,$^{3,4}$
Massimo Dotti,$^{3,4}$
Melvyn B. Davies$^{5}$
and Piero Madau$^{6}$
\\
$^{1}$Center for Theoretical Astrophysics and Cosmology, Institute for Computational Science, University of Zurich, Winterthurerstrasse\\ 190, CH-8057 Z\"urich, Switzerland\\
$^{2}$Scuola Normale Superiore, Piazza dei Cavalieri 7, IT-56126 Pisa, Italy\\
$^{3}$Dipartimento di Fisica ``G.~Occhialini'', Universit\`{a} degli Studi di Milano-Bicocca, Piazza della Scienza 3, IT-20126 Milano, Italy\\
$^{4}$INFN, Sezione di Milano-Bicocca, Piazza della Scienza 3, IT-20126 Milano, Italy\\
$^{5}$Lund Observatory, Department of Astronomy and Theoretical Physics, Lund University, Box 43, SE-221 00 Lund, Sweden\\
$^{6}$Department of Astronomy and Astrophysics, University of California at Santa Cruz, 1156 High St., Santa Cruz, CA 95064, USA
}

\date{Accepted XXX. Received YYY; in original form ZZZ}

\pubyear{2020}

\begin{document}
\label{firstpage}
\pagerange{\pageref{firstpage}--\pageref{lastpage}}
\maketitle

\begin{abstract}
The forthcoming Laser Interferometer Space Antenna (LISA) will probe the population of coalescing massive black hole (MBH) binaries up to the onset of structure formation. Here we simulate the galactic-scale pairing of $\sim$10$^6$~M$_{\sun{}}$ MBHs in a typical, non-clumpy main-sequence galaxy embedded in a cosmological environment at $z=7$--6. In order to increase our statistical sample, we adopt a strategy that allows us to follow the evolution of six secondary MBHs concomitantly. We find that the magnitude of the dynamical-friction induced torques is significantly smaller than that of the large-scale, stochastic gravitational torques arising from the perturbed and morphologically evolving galactic disc, suggesting that the standard dynamical friction treatment is inadequate for realistic galaxies at high redshift. The dynamical evolution of MBHs is very stochastic, and a variation in the initial orbital phase can lead to a drastically different time-scale for the inspiral. Most remarkably, the development of a galactic bar in the host system either significantly accelerates the inspiral by dragging a secondary MBH into the centre, or ultimately hinders the orbital decay by scattering the MBH in the galaxy outskirts. The latter occurs more rarely, suggesting that galactic bars overall promote MBH inspiral and binary coalescence. The orbital decay time can be an order of magnitude shorter than what would be predicted relying on dynamical friction alone. The stochasticity, and the important role of global torques, have crucial implications for the rates of MBH coalescences in the early Universe: both have to be accounted for when making predictions for the upcoming LISA observatory.
\end{abstract}




\section{Introduction}\label{sec:intro}

The coalescence of massive black hole (MBH) binaries will constitute one of the loudest gravitational-wave (GW) events in the low-frequency band of the forthcoming space-borne observatory Laser Interferometer Space Antenna \citep[LISA;][]{Amaro-Seoane2017,Barack_et_al_2019}. LISA will detect mergers between $10^4$--$10^7$~M$_{\sun{}}$ MBHs up to and above $z\sim20$, thus probing the galaxy and MBH clustering up to the onset of structure formation.

The evolution of MBH pairs formed as a result of a galaxy merger has been first pictured in the pioneering work by \citet{Begelman1980}: 
in simple terms, dynamical friction \citep[DF;][]{Chandrasekhar1943} brings the MBHs down to the scale at which they form a bound system; %
shortly afterwards, the binary energy and angular momentum get drained via repeated three-body scatterings with interacting stars \citep[e.g.][]{Saslaw1974} or via its interaction with gas; finally, at roughly mpc scales, GWs start dominating the evolution and lead to coalescence \citep[e.g.][]{Thorne1976}. \\

The details of this rather simple picture have been largely investigated in the past few decades. A number of pieces of literature focused on the relatively small scale evolution (100--0.01~pc), and in particular on the effects of gas \citep[e.g.][]{Armitage2002, Escala2005, Dotti2007, Mayer2007, Goicovic2017, Souza-Lima2020} and on the efficiency of three-body stellar scatterings in ensuring the GW phase is eventually reached \citep[e.g.][]{Milosavljevic2003, Khan2011,  Vasiliev2015, Bortolas2016, Rantala2017, Gualandris2017, Rantala2018, Bortolas2018, Bortolas2018tr,Mannerkoski2019}.
The majority of these studies hint to an effective inspiral  in both stellar and gaseous environments, with characteristic time-scales  ranging from $\sim$10~Myr to a few Gyr, depending on the properties of the background environment. 

The larger-scale, DF-driven inspiral phase has been long regarded as relatively easy to model: a vast number of works confirmed to a good extent the well-known \citeauthor{Chandrasekhar1943} predictions for the decay time-scale \citep[e.g.][]{Boylan-Kolchin_et_al_2008}, 
highlighting a prompt inspiral for galaxy collisions with mass ratios above 1:5 \citep[e.g.][]{Taffoni2003}, and a less efficient pairing otherwise, as the DF-induced deceleration scales with the inverse of the perturber's mass. Moreover, a smaller intruder galaxy is likely to get severely tidally- and ram-pressure-stripped prior to reaching the main galaxy's interiors \citep[e.g.][]{Callegari2011,VanWassenhove_et_al_2014,Capelo_Dotti_2017}. Deviations from the standard DF behaviour have been reported mainly if the (non-gas-dominated) host galaxy features a significantly cored profile\footnote{Gas-poor galaxies which previously experience an MBH coalescence could feature extremely extended cores, if the GW recoil associated to the coalescence significantly displaced the new-born MBH, whose subsequent inspiral injected energy in the surrounding environment \citep{Gualandris2008}.} \citep[e.g.][]{Bonfini2016}: in such systems, DF efficiency can be remarkably suppressed \citep{Read2006, Tamfal2018}, and one has to account for both stars moving slower and faster than the perturber to properly model the orbital decay \citep{Antonini2012b}.

Note  that  nearly all studies  addressing the large-scale pairing of two MBHs have been carried out in very idealized frameworks: either two MBHs are placed in an equilibrium (galaxy) model \citep[e.g.][]{Gualandris2017,Tamburello2017}, or two galaxies and their embedded MBHs are allowed to come together in isolation \citep[e.g.][]{Capelo_et_al_2015, Bortolas2018tr}. This idealized approach has the advantage of requiring reasonable computational resources while allowing to study the MBH pairing down to sub-kpc and sometimes even sub-pc scale \citep[e.g.][]{Khan2018}. On the other hand, this strategy discards the cosmological context and its implications, e.g. the possibility of distorted, non-symmetric galaxy morphologies, galaxy fly-bys, morphological transformations, and so forth. Accounting for the larger-scale environment is important when addressing the MBHs' path to coalescence in the present-day Universe, and  becomes crucial in the infant Universe that LISA is going to probe. 

In fact, even in the absence of a complete cosmological scenario, a series of studies on the pairing of MBHs in star-forming galaxies at $z \sim1$--2 highlighted that the formation of numerous massive star-forming clumps (with masses larger than the MBH by orders of magnitude) within the galaxy could scatter the inspiralling MBH orbit and render its evolution  stochastic at the scale of the circumnuclear disc \citep{Fiacconi2013, del-Valle2015, Souza-Lima2017} or even the entire galactic disc \citep{Roskar2015, Tamburello2017}.

\citet{Khan2016} were the first to address the MBH pairing from cosmological distances all the way down to the hardening and coalescence phase. However, they focused on a very massive galaxy  at  $z\approx3$  [this would become a  brightest cluster galaxy (BCG) by $z=0$, see \citealt{Feldmann2015}]; this relatively
rare object, which underwent quenching after a merger-driven starburst, rapidly exhausted its gas reservoir, allowing the authors
to turn the calculation into a pure $N$-body one, thus discarding the role of hydrodynamics. They found a surprisingly short coalescence time-scale ($\sim$10~Myr), about two orders of magnitude smaller than what was previously obtained in idealized merger simulations at $z \approx 0$ \citep[e.g.][]{Khan2011}.

More recently, attempts to account for the cosmological framework including  hydrodynamics 
started to come forward \citep[e.g.][]{Tremmel2018, Tremmel2018b, Pfister2019}. These studies hint to a more complex evolution compared to what would be predicted via a simple application of the theory of DF \citep[e.g.][]{Chandrasekhar1943,Ostriker1999,Colpi1999}
and suggest that the time-scale for the pairing in these more realistic environments is much less predictable, especially for the relatively low-mass (<$10^5$~M$_{\sun{}}$) MBHs that experience a weaker DF \citep[e.g.][]{Bellovary_et_al_2019}. 
However, the mass and spatial resolution limits in the aforementioned  studies 
often require to introduce DF as an external drag force in order to properly follow the 
orbital decay of MBHs at sub-galactic scales.

Motivated by these considerations, here we follow the kpc-scale pairing of LISA MBHs in a typical, non-clumpy main-sequence star-forming galaxy at $z = 7$--6 using unprecedented mass and force resolution. 

\section{Methods}\label{sec:methods}


We study the evolution of the pairing of MBHs embedded in a star-forming main-sequence galaxy at $z \lesssim 7$. Our initial conditions are taken from the $z = 7.32$ snapshot of the high-resolution cosmological zoom-in simulation PonosHydro (PH; presented in \citealt{Fiacconi2017}) -- the precursor run -- to which we add the MBHs, as described in detail in Section~\ref{sec:BH_initialization}. We run our simulation using the smoothed particle hydrodynamics, $N$-body code {\textsc{gasoline2}}\footnote{The integrator employs a KD-tree scheme for solving gravity and a leapfrog scheme for handling the time integration. Thus, {\textsc{gasoline2}} is well suited for studying the large scale decay of MBHs, but is inadequate to follow the subsequent stellar hardening phase that follows the binary formation \citep{Gualandris2017}. Here we do not attempt to follow the latter.  } \citep{Stadel_2001,Wadsley2004,Wadsley2017}, using a setup that is virtually identical to the parameters set for the PH run in \citet{Fiacconi2017}. The only different parameter between our runs and the one presented in \citet{Fiacconi2017} is the maximum allowed timestep, which in our case is $\approx 0.2$~Myr, whereas it was $\approx 16$~Myr in the precursor run. This choice allows us to obtain a fine-grained collection of snapshots, from which we can derive the relevant quantities for the analysis with an adequate time resolution.

In the following sections, we describe the precursor and current runs. Unless otherwise stated, all quantities are expressed in physical units, the time coordinate has to be intended as the time since the Big Bang, and the adopted $\Lambda$-CDM cosmology is consistent with the results of the \textit{Wilkinson Microwave Anisotropy Probe} 7/9 years ($\Omega_{\rm m,0} = 0.272$, $\Omega_{\rm \Lambda,0} = 0.728$, $\Omega_{\rm b,0} = 0.0455$, $\sigma_8 = 0.807$, $n_{\rm s} = 0.961$, and $H_0 = 70.2$~km~s$^{-1}$~Mpc$^{-1}$; \citealt{Komatsu2011, Hinshaw2013}).


\subsection{Precursor run}\label{sec:precursor}


PH is a high-resolution, hydrodynamics, $N$-body version of the dark matter-only zoom-in cosmological simulation PonosV from the Ponos suite by \citet{Fiacconi2016}, which was run in a box of 85.5~cMpc using the code {\textsc{pkdgrav3}} \citep[][]{Potter_et_al_2017}.

The initial mass of gas and (high-resolution) dark matter particles in PH is 883.4 and 4397.6~M$_{\sun{}}$, respectively. The physical gravitational softening of those particles is, for $z \le 9$, 47.9~pc for baryons and 81.8~pc for dark matter.\footnote{For $z > 9$, these quantities need to be multiplied by $1/(1+z)$.}

In the precursor run, gas can cool radiatively, via both non-equilibrium cooling of atomic primordial species and photoionization equilibrium cooling of metal species \citep[][]{Shen_et_al_2013}, assuming a redshift-dependent photoionizing UV background \citep{Haardt_Madau_2012}.

Star formation and stellar feedback are modelled following the recipes of \citet{Stinson_et_al_2006}. More specifically, gas particles are allowed to form stars only if the local gas density and overdensity are higher than $10 \,m_{\rm H}$~cm$^{-3}$ and 2.64, respectively, the local gas temperature is lower than $10^4$~K, and the flow is convergent and locally Jeans-unstable. If all these criteria are met, then gas particles are stochastically selected such that, on average, ${\rm d}M_*/{\rm d}t = \epsilon_* M_{\rm gas} /t_{\rm dyn}$, where $M_*$ and $M_{\rm gas}$ are the mass of stars and gas involved, respectively, $\epsilon_{*} = 0.05$ is the star formation efficiency, and $t_{\rm dyn}$ is the local dynamical time. The newly formed stellar particles have a mass of $353.4$~M$_{\sun{}}$: being much more massive than real individual stars, they represent a stellar population, which is described by a \citet{Kroupa2001} initial mass function.

Stars with masses between 1 and 8~M$_{\sun{}}$ release part of their mass as stellar winds, with the returned gas retaining the metallicity of those stars. Stars with masses between 8 and 40~M$_{\sun{}}$, depending on their mass-dependent lifetime \citep[][]{Raiteri_et_al_1996}, can explode as Type~II supernovae: they inject into the surrounding gas a given amount of total, iron, and oxygen mass, dependent on the progenitor stellar mass \citep[][]{Woosley_Weaver_1995,Raiteri_et_al_1996}, and energy ($10^{51}$~erg), according to the `blastwave model' of \citet{Stinson_et_al_2006}, in which cooling is temporarily disabled. Type~Ia supernovae are also modelled and inject $10^{51}$~erg and a fixed amount of mass and metals \citep[][]{Thielemann_et_al_1986,Raiteri_et_al_1996}.

In addition to the standard SPH formulation, we include thermal energy and metal turbulent diffusion \citep[with a diffusion coefficient $C = 0.05$;][]{Wadsley_et_al_2008,Shen_et_al_2010}, and a pressure floor \citep[][]{Roskar2015}, to have a minimum number of resolution elements that must resolve the Jeans length and ensure the correct fragmentation behaviour of the gas. The gas becomes therefore non-ideal when the pressure floor is reached, since the temperature can cool below the ideal-gas equivalent of the pressure floor \citep[see, e.g.][]{Capelo_et_al_2018}.

For more details about the PH simulation and its dark matter-only counterpart (PonosV), we refer the reader to \citet{Fiacconi2017} and \citet{Fiacconi2016}, respectively.


\subsection{Current run}\label{sec:current}


\begin{figure}
\centering
\includegraphics[ width=.45\textwidth]{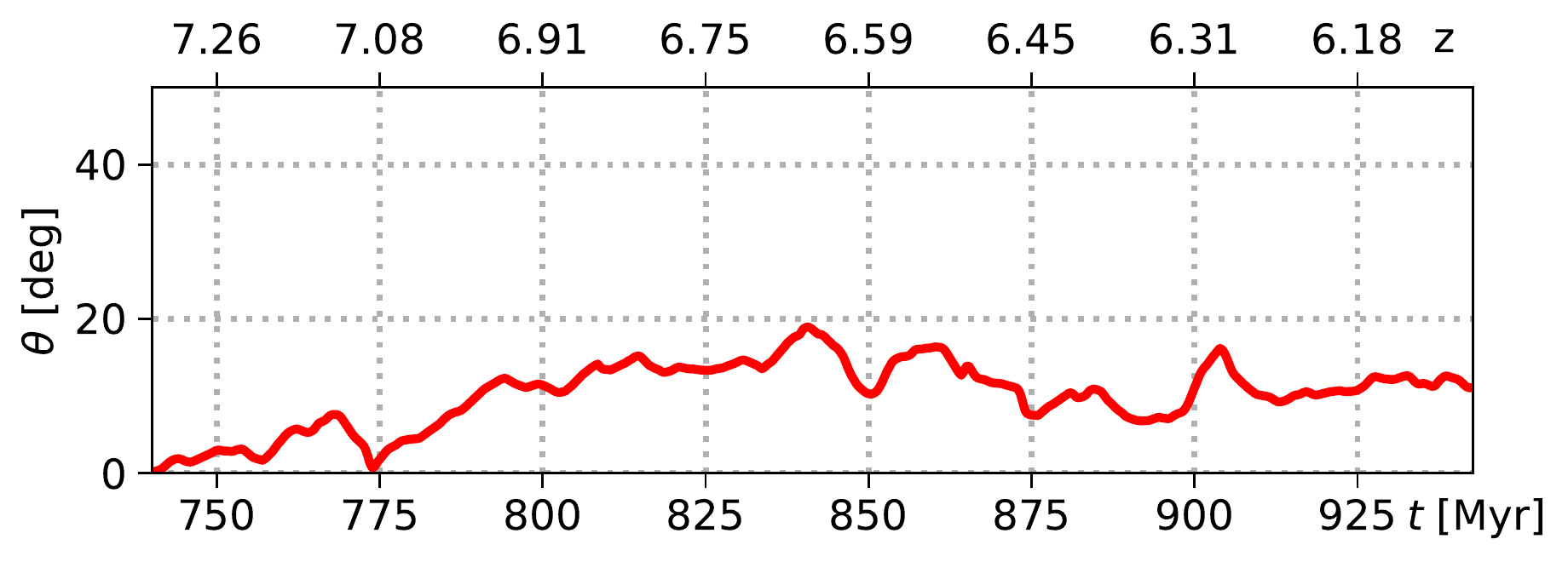}
    \caption{Angle $\theta$ between the gas angular momentum within the central 0.5~kpc and its value in the initial conditions as a function of time. 
    The orientation of the gas angular momentum within this radius is used as a proxy for the galaxy orientation throughout the paper.}
    \label{fig:Global_am_oscillation}
\end{figure}

\begin{figure}
\centering
\includegraphics[width=0.45\textwidth]{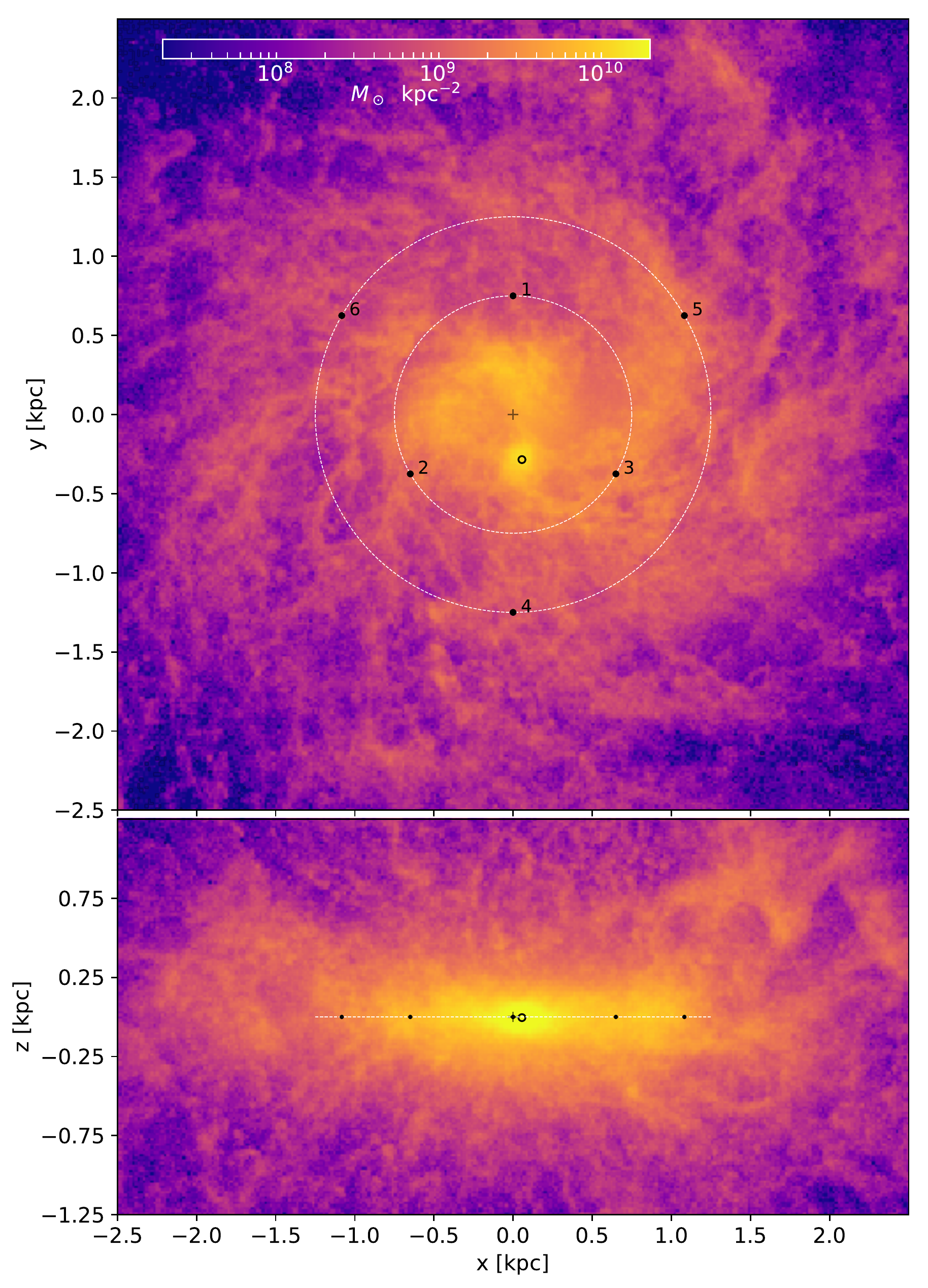} 
    \caption{Face-on (top) and edge-on (bottom) view of the galaxy's baryonic projected mass density at the beginning of the integration ($z\approx7.3$). The secondary MBHs are represented with labelled  black dots, whereas the primary is shown as an empty circle and the centre of the galaxy (defined in Section~\ref{sec:BH_initialization}) is marked with a cross. The white thin lines trace the expected initial orbits for the secondary MBHs, that are initialized on circular trajectories that move in the clockwise direction in the top panel. }
    \label{fig:IC_circ}
\end{figure}

The  simulation presented in this paper is relatively expensive and the human time needed for the integration can get to several months even with the best current present-day facilities. On the other hand, the result of a single run  has to be taken with caution, as its statistical significance cannot be addressed. 

For these reasons, here we adopt  a workaround that allows us to study the pairing of multiple MBH pairs within a single run: we place six secondary MBHs in the main galaxy and we study the pairing of each of those with the (unique) primary MBH. Our idea is justified by the different mass-scales in our system:

\begin{equation}
    m_{\rm particle} \ll M_{\rm MBH} \ll  M_{\rm system},
\end{equation}

\noindent that is, the mass of the typical particle in our system ($ m_{\rm particle}\lesssim 4000$~M$_{\sun{}}$ in the high-resolution region) is much smaller than the mass we assign to the MBHs ($\sim$10$^6 $~M$_{\sun{}}$) which in turn is much smaller than the mass of the host galaxy ($M_{\rm system}\approx 5\times 10^{9} $~M$_{\sun{}}$ within the central 1.25~kpc in the initial conditions). For this, unless the MBHs reach distances small enough to scatter each other, we can assume them to evolve independently. In Section~\ref{sec:MBHscattering}, we describe in detail how we check for spurious scatterings.

\subsubsection{Initialization of the MBHs}\label{sec:BH_initialization}

In what follows, we refer to the {`centre of the galaxy'} as the centre of mass of all particles (excluding the MBHs) in the galaxy's innermost 0.5~kpc, iteratively recomputed until a tolerance of 0.1~pc is achieved; the velocity of the galaxy is defined as the mass-weighted velocity of the same particles. Analogously, the galaxy angular momentum (and galaxy plane) is obtained by computing the gas angular momentum within the central 0.5~kpc. These quantities are re-computed for each snapshot in the simulation and used as 
the reference frame for the analysis; in particular, we choose the $z$ axis to be parallel to the galaxy angular momentum. For reference, Figure~\ref{fig:Global_am_oscillation} shows the evolution of the angle between the galaxy angular momentum and the same quantity computed at the beginning of the integration. This angle never exceeds 20 degrees, suggesting that the galaxy plane does not significantly oscillate about its initial position.

The presented reference frame is also adopted for initializing the secondary MBHs. In fact, these objects respond to the bulk motion of the galaxy. We set the six secondary MBHs in two triangular configurations as shown in Figure~\ref{fig:IC_circ}: three of them (labelled as 1--3) have an initial separation from the centre of 0.75~kpc, whereas the other three (labelled as 4--6) are placed at a radius of 1.25~kpc. The MBHs initially have a tangential velocity equal to the velocity of a circular orbit given by the enclosed mass within their radius as if the galaxy were spherically symmetric: such velocity is equal to 116 and 128~km~s$^{-1}$ for MBHs initially at 0.75 and 1.25~kpc from the centre, respectively. The initial angular momentum direction of the MBHs coincides with the galactic one.

The primary MBH is not placed at the galaxy centre but in the densest region near it, found via the shrinking sphere mechanism, 
as this is a more likely region for its formation;  
its initial velocity is equal to the centre of mass velocity of particles in a sphere of 0.2~kpc around it. Note that the primary MBH and the centre of the galaxy are initially separated by a distance of 0.29~kpc, which gradually shrinks to zero in $\sim$200~Myr (the evolution of this separation in the run is shown in Figure~\ref{fig:Separation}).

The primary MBH has a mass $M_1 = 4\times 10^6$~M$_{\sun{}}$ 
i.e. 1/500 of the total mass in stars within the galaxy virial radius, in relatively good agreement with what suggested in \citet{Trakhtenbrot2010} and \citet{Lupi2019}. The secondaries are assigned a mass $M_2 = 10^6$~M$_{\sun{}}$; they are inserted in the simulation without any host galaxy remnant, in the assumption that all the mass of the parent system that brought them into the main halo was stripped. Such stripping event is more likely for mergers with mass ratio $\lesssim$1:10 \citep[e.g.][]{Callegari2011,VanWassenhove_et_al_2014}. However, we keep the mass ratio between the primary and the secondaries $M_2/M_1=0.25$, as this could mimic the presence of a stellar compact nucleus around the secondary,  that survived stripping; in addition, a large $M_2$ enhances the effect of DF that might be not completely resolved in our runs, as also mentioned in Section~\ref{sec:directcompDFtorque}.

The MBHs have a softening equal to 10 pc. MBH accretion and feedback are switched off in the present run, as this allows us to better control the effect of galactic structures on the orbital evolution, net of radiative effects; note that MBH feedback is suspected to have a small effect on the overall galactic structure if the MBH mass is smaller than $\sim$10$^7$~M$_{\sun{}}$ \citep[e.g.][]{Angles-Alcazar2017}.

\subsubsection{Check for MBH scatterings}\label{sec:MBHscattering}

The mass enclosed within a sphere of 0.1~kpc around each secondary MBH at the beginning of the integration is larger than $2M_2$, implying that the MBHs need to get closer than 0.1~kpc to scatter each other. Whenever this happens, we find the smallest distance attained by the two MBHs and compute the mass enclosed within a sphere of radius equal to their minimum distance (excluding the mass of the MBHs themselves). If this mass is smaller than $2 M_2$, we take one of the two MBHs out of the simulation and we restart it before the scattering occurred.

Spurious scattering occurs only once in the run: MBHs 1 and 5 encounter each other at $t\approx$ 797~Myr; the former is thus removed. 
MBHs 3 and 4 also get very close to each other at $t \approx 795$ Myr, 
but they do not meet the conditions for one of them to be excluded.

Note that, whenever a secondary gets close enough to the primary MBH, it enters a regime in which we are unable to resolve the subsequent, small-scale inspiral. In particular, we consider the large-scale inspiral to be completed when the periapsis reaches 0.1~kpc; we remove the MBH shortly afterwards. The orbit of the primary is not going to be strongly affected by the inspiralling secondaries, as there is a dense nucleus around it whose mass at the closest approach with each secondary is always larger than $4\times 10^6 $~M$_{\sun{}}$.

Another source of spurious scattering could be constituted by contaminating (low-resolution) dark matter particles; we checked that the contaminant fraction is  always zero within the innermost 15~kpc, i.e. within a radius much larger than our region of interest.

\section{Results}\label{sec:results}

The properties of the main galaxy are broadly detailed in \citet{Fiacconi2017}: the galaxy harbours gas in a hot and turbulent state, and it lacks massive star forming clumps. It displays an evolving, geometrically thick stellar and gaseous disc whose typical stellar (gaseous) scale-length, scale-height, and total mass are, respectively, about 0.5~kpc, 0.2~kpc, and $5 \times 10^8$~M$_{\sun{}}$ (0.8~kpc, 0.3~kpc, and $\sim5 \times 10^8$~M$_{\sun{}}$), whereas the galaxy's virial mass is $\sim 10^{11}$~M$_{\sun{}}$; these numbers are estimated at the beginning of the integration but can be referenced at least on an order-of-magnitude level for the whole integration. 

\subsection{Black hole orbital evolution} \label{sec:orb_ev}

\begin{figure*}
\centering
\includegraphics[ width=.99\textwidth]{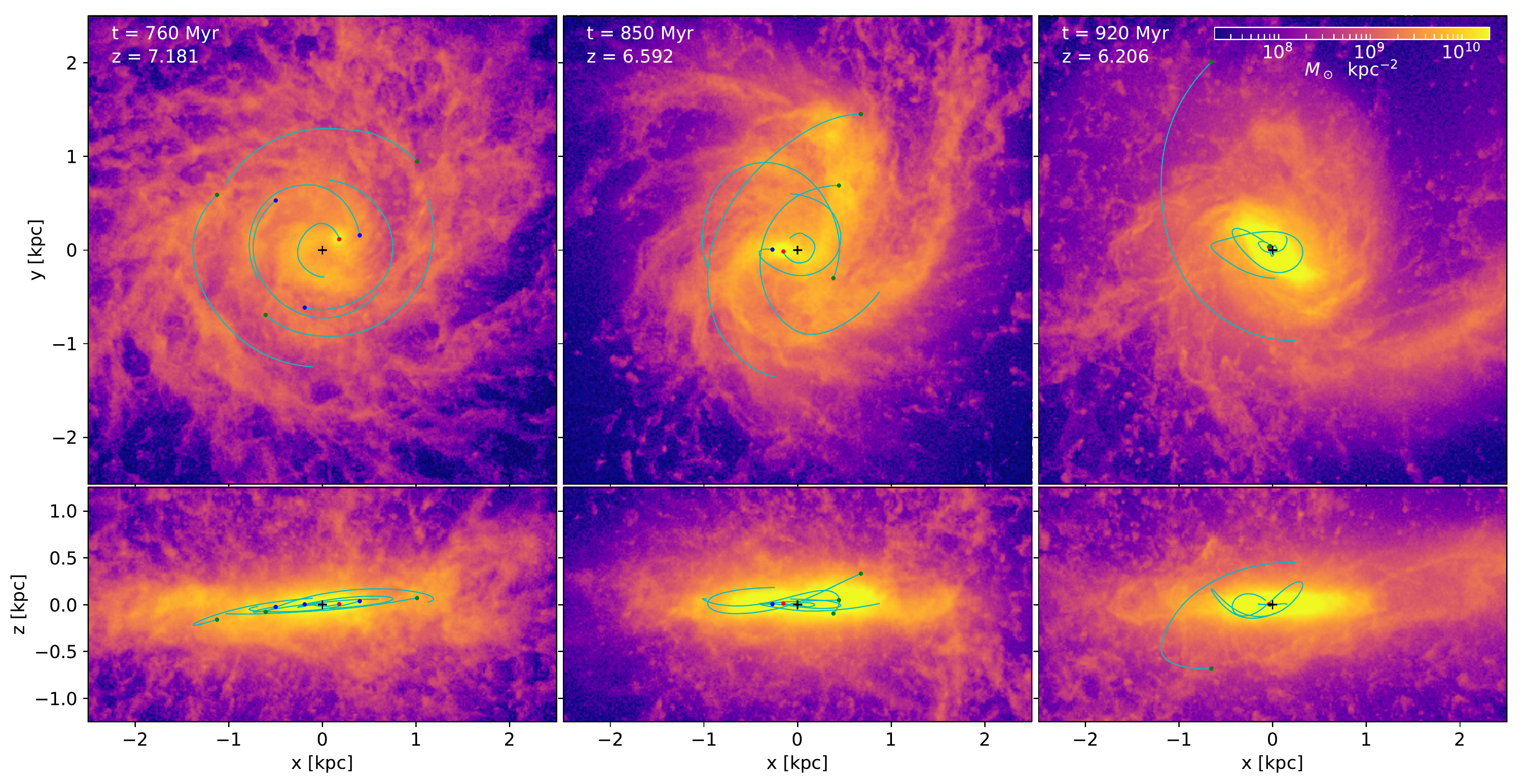}
    \caption{
    Face-on (top panels) and edge-on (bottom panels) baryonic surface density at three different times as labelled (see columns). The positions of the MBHs are shown with a small coloured dot: red for MBH~0 (the primary), blue for MBHs 1--3 (initially at 0.75~kpc from the centre), and green for MBHs 4--6 (initially at 1.25~kpc from the centre). The cyan trajectories show the orbit of the MBHs in the 20~Myr prior to the time of each snapshot. The black cross marks the centre of the galaxy.
    }
    \label{fig:snapshots}
\end{figure*}

\begin{figure*}
\centering
\includegraphics[ width=.75\textwidth]{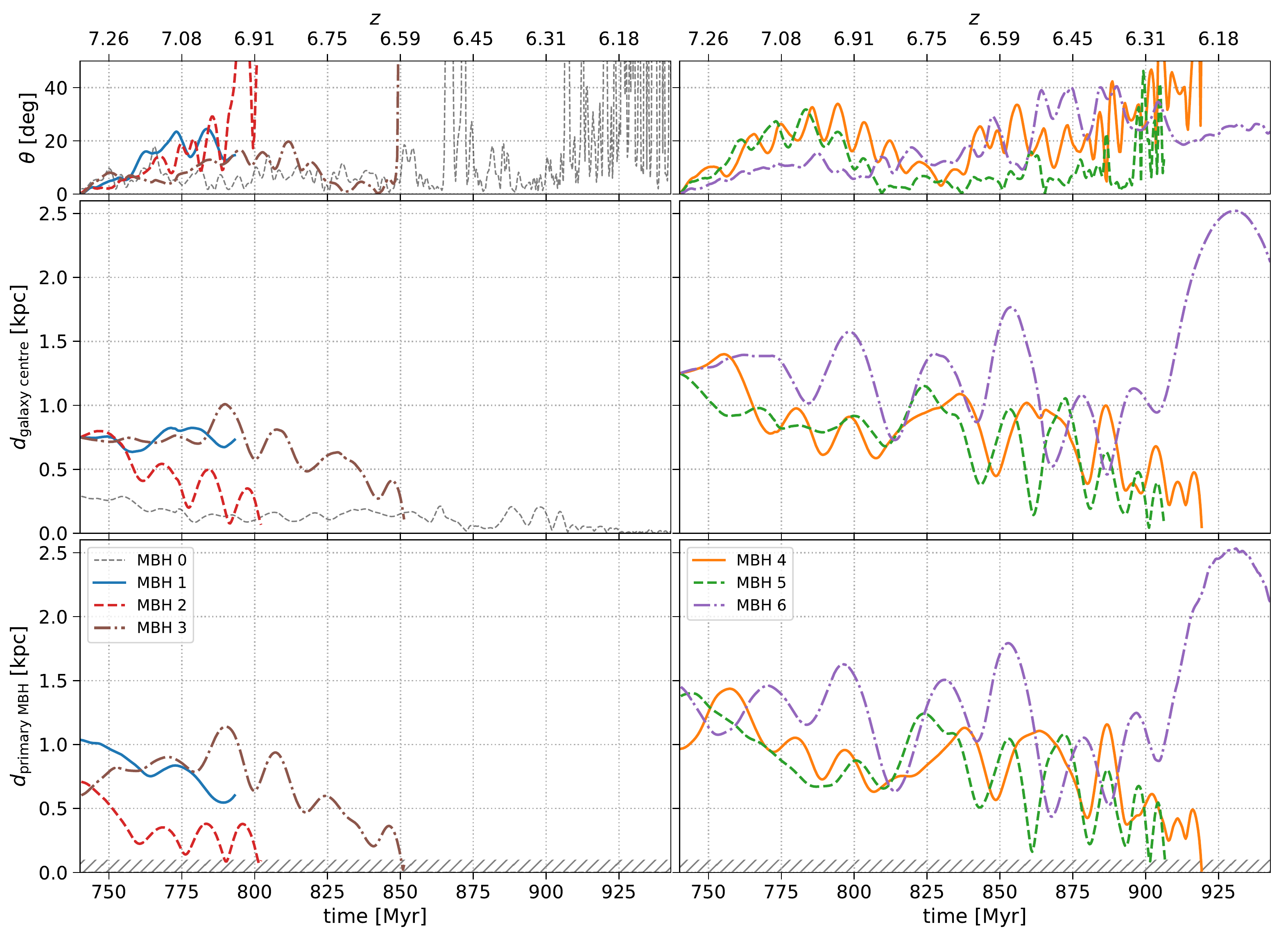}
    \caption{Orbital evolution of the MBHs as a function of time. 
    The left-hand panels refer to the primary MBH (MBH~0) and to the MBHs with an initial separation from the centre of 0.75~kpc (MBHs~1--3), whereas the right-hand panels show the evolution of the MBHs initially displaced 1.25~kpc from the centre (MBHs 4--6). The top row shows the angle between the MBHs orbital plane and the galaxy plane; the central row shows the separation between each MBH and the galaxy centre; and the bottom row shows the distance of each secondary MBH from the primary. The hatched area in the bottom panels marks the region in which the orbital evolution of the secondaries cannot be resolved in our run, as their separation from the primary drops below 0.1~kpc. }
    \label{fig:Separation}
\end{figure*}

Figure~\ref{fig:snapshots} shows the position of the MBHs in the galaxy in three different snapshots of the simulation, whereas Figure~\ref{fig:Separation} shows their orbital evolution as a function of time. 

The primary MBH initially oscillates around the centre of the galaxy, as its orbital decay is not completed yet. In fact, the nuclear overdensity in which the primary has been placed is the remnant of a lighter galaxy that merged with the main galaxy some $\approx 50$~Myr before the start of the present simulation. The intruder's dense nucleus overtook the nucleus of the main galaxy in a \textit{nuclear coup} fashion \citep{VanWassenhove_et_al_2014}, and it is on its way to the centre of the galaxy.

We now focus on the orbital-decay time of the secondary MBHs (last column of Table~\ref{tab:timescale}). 
Since MBH~1 is prematurely removed from the simulation, we can only conclude that its inspiral time would have been longer than $\approx 60$ Myr. MBHs~2 and 3 take $\approx 50$ and $\approx 110$~Myr, respectively, to reach a separation smaller than 0.1~kpc from the primary. This factor >2 in the inspiral times suggests a certain degree of stochasticity in the orbital evolution, especially considering that both MBHs started at the same radius (0.75~kpc). 

The effect of stochasticity is even more prominent in the orbital evolution of MBHs with an initial separation of 1.25~kpc from the centre: among those, MBHs~4 and 5 complete their inspiral, respectively, in 140 and 160~Myr (i.e. their inspiral time differs by only $\sim$15 per cent); on the other hand, MBH 6 never attains a separation smaller than 0.5~kpc from the centre, and its final radius (and angular momentum magnitude) is doubled if compared to the initial one; its inspiral time is likely going to be much longer  than the simulated time of $\approx 210$~Myr. We anticipate here that the overall orbital evolution of MBHs~4--6 (and perhaps even MBH~3) is significantly affected by the development of a strong bar whose dynamical effect can be inferred already at time $\approx 840$ Myr, as discussed below; the presence of the bar becomes obvious even by eye-inspection after $t = 880$ Myr (see Figure~\ref{fig:snapshots}).

\subsection{Dynamical friction versus gravitational torques}

In this section, we address the main mechanisms responsible for the MBHs orbital evolution. In particular, we compare the global gravitational torques\footnote{In what follows, torques and angular momentum are expressed per unit mass (reduced quantities).} acting on each MBH to the DF-induced torques, whereas the effect of the bar alone is detailed in Section~\ref{sec:bar}. %
\textit{DF-induced torques can be thought of as the torques originated by the response of the host to the passage of the MBHs. Global torques, instead, are induced by all non-symmetries  and inhomogeneities in the system regardless of their origin.}\footnote{Note that the two are heavily interconnected: in fact, the non-DF-induced inhomogeneities that scatter the orbit of the MBHs are the same that randomize the effectiveness of DF, as better discussed in Section~\ref{sec:discussion}. }

The global gravitational torques acting on each MBH are obtained by computing the total gravitational acceleration $\mathbf{a}_{\rm tot}$ due to {\it all} particles in the system,\footnote{The global acceleration $\mathbf{a}_{\rm tot}$ was extracted via the KD-tree implemented in {\textsc{pkdgrav3}} \citep[][]{Potter_et_al_2017}.} including particles in the immediate vicinity of the MBH; this implies that $\mathbf{a}_{\rm tot}$ contains the component of the acceleration associated to DF. The global torque acting on each MBH is then computed as ${\boldsymbol\tau}_{\rm tot} = \mathbf{r}\times\mathbf{a}_{\rm tot}$, where $\mathbf{r}$ is the MBH position vector and~$\times$ is the cross product. 

The DF-induced torque, instead, is obtained via a semi-analytical approach related to the DF modelling by \citet{Chandrasekhar1943} and \citet{Ostriker1999}, as detailed below.

\subsubsection{Dynamical friction estimate}

\begin{figure*}
\centering
\includegraphics[ width=.99\textwidth]{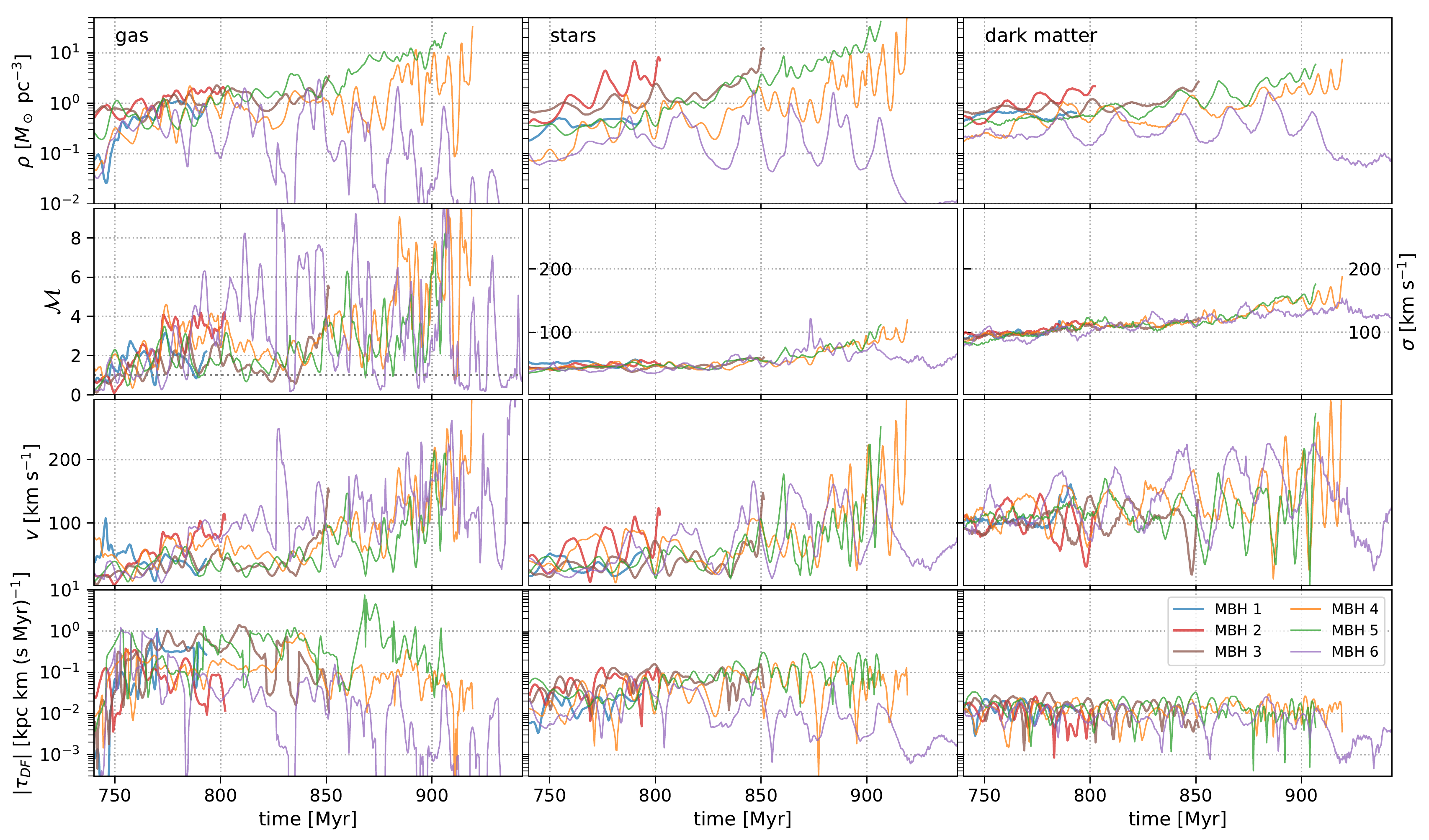}
    \caption{Temporal evolution of the relevant quantities used for the computation of DF acting on each secondary MBH. The first, second, and third column respectively refer to quantities associated to the gaseous, stellar, and dark matter background, respectively. The top row shows the evolution of the local density around each MBH. 
    The second row displays the Mach number (for gas) and the velocity dispersion (for stars and dark matter): note the different $y$ scale. The third row shows the relative velocity between each MBH and the background medium. The fourth row shows the modulus of the DF-induced torque due to each of the different components. All relevant quantities are computed within a sphere of 0.15~kpc around each MBH.
    }
    \label{fig:DF_components}
\end{figure*}

Our simulation features three different components that contribute to DF, i.e. gas, stars, and dark matter. The gas-induced deceleration can be estimated as \citep{Ostriker1999}

\begin{equation}
    \frac{{\rm d} \mathbf{v}_{\rm g}}{{\rm d}t}\Big\rvert_{\rm DF}   = -4 \pi \ln \left[ \frac{b_{\rm max}}{b_{\rm min}} \frac{(\mathcal{M}^2-1)^{1/2}}{\mathcal{M}}\right]G^2 M_2 \rho_{\rm g} \frac{\mathbf{v}_{\rm g}}{\left|\mathbf{v}_{\rm g}\right|^3},
    \label{eq:df_gas}
\end{equation}

\noindent where $\rho_{\rm g}$ is the gas density, $\mathcal{M} = |\mathbf{v}_{\rm g}|/c_{\rm s}$ is the Mach number, $c_{\rm s}$ is the speed of sound, $\mathbf{v}_{\rm g}$ is the local relative velocity between the MBH and the surrounding gas, and $G$ is the gravitational constant. Note that this prescription is valid only if $\mathcal{M}>1$, whereas, for subsonic motion, we replace the natural logarithm in Equation~\eqref{eq:df_gas} with $\mathcal{M}^3/3$. The value of the minimum  and maximum impact parameters, $b_{\rm min}$ and  $b_{\rm max}$, are discussed below.

The DF-induced deceleration due to the two non-dissipative components, i.e. stars and dark matter, can be written as \citep{Chandrasekhar1943}\footnote{This approach implicitly assumes a Maxwellian distribution for the velocity of stars and neglects the contribution of stars moving faster than the MBH. Nonetheless, we expect our approach to give a good estimate for DF, considering that the gas component is the main responsible for DF.}

\begin{equation}
\begin{split}
    \frac{{\rm d} \mathbf{v}_x}{{\rm d}t}\Big\rvert_{\rm DF}   =&
    -2 \pi \ln \left( 1+\frac{b^2_{\rm max}}{b^2_{\rm min}}\right) G^2 M_2 \rho_x \ \frac{\mathbf{v}_x}{\left|\mathbf{v}_x\right|^3} \times  \\
    &\left[ {\rm erf} \left(\frac{\left|\mathbf{v}_x\right|}{\sqrt{2}\sigma_x} \right) - \left( \sqrt{\frac{2}{\pi}} \frac{\left|\mathbf{v}_x\right|}{\sigma_x} \right) \exp \left(-\frac{\left|\mathbf{v}_x\right|^2}{2\sigma_x^2}\right)\right],
\end{split}    
    \label{eq:df_stars}
\end{equation}

\noindent where the subscript $x$ 
refers to quantities computed with respect to the stellar or dark matter background, $\rho_x$ is the associated local density, $\mathbf{v}_x$ is the relative velocity between the MBH and the surrounding medium $x$, and $\sigma_x$ is the one-dimensional velocity dispersion.

The value of the maximum impact parameter, $b_{\rm max}$, is set equal to 0.6~kpc for the stellar and gaseous component; this value corresponds to a few disc scale-heights, as the underlying assumption of quasi-homogeneous density fails at larger scales. The value of $b_{\rm max}$ is chosen to be 3~kpc for dark matter, i.e.  of the order of the scale radius of a \citet{Navarro1996} profile whose fit was performed at the beginning of the integration. For each component, the minimum impact parameter is computed as

\begin{equation}\label{eq:bmin}
b_{\rm min} = \dfrac{GM_2}{\sigma_{bg}^2},
\end{equation}

\noindent where $\sigma_{bg}$ is the velocity dispersion of each background species.

In order to constrain the effects of DF, we compute the relevant quantities in Equations~\eqref{eq:df_gas}--\eqref{eq:bmin} within a sphere of radius $R = 0.15$~kpc around each secondary MBH. The Mach number is obtained by mass-averaging the speed of sound of all particles in the sphere. This quantity is computed as $c_{\rm s} = \sqrt{\gamma_a k_{\rm B} T/(\mu m_{\rm p})}$, where $T$ is the gas temperature, $\gamma_a=5/3$ is the adiabatic index, $k_{\rm B}$ is the Boltzmann constant, $\mu \approx $ 0.6 is the mean molecular weight, and $m_{\rm p}$ is the proton mass. 
The relative velocity ($\mathbf{v}_{\rm g}$, $\mathbf{v}_x$)
is obtained as the mass-weighted mean of the relative velocity between the MBH and the particles of each background component within the sphere; the associated velocity dispersion is estimated as $\sqrt{\sum_i \sigma_i^2/3}$, where $i$ runs over the three spatial directions.   

Finally, the DF-induced torque can be obtained via

\begin{equation}
{\boldsymbol\tau}_{{\rm DF}, bg} = \mathbf{r} \times \frac{{\rm d}\mathbf{v}_{bg}}{{\rm d}t}\Big\rvert_{\rm DF}   
\end{equation}

\noindent for each component $bg$ of the background (gas, stars, and dark matter), where $\mathbf{r}$ is again the MBH position vector. 
In order to assess the validity of our computation, we checked that the DF-torque given by the sum of the three background components does not change significantly (i) when varying the radius $R$ of the sphere over which the relevant quantities in Equations~\eqref{eq:df_gas}--\eqref{eq:df_stars} are computed ($R = 0.05$, 0.10, 0.15, and 0.2~kpc); (ii) when fixing the Mach number to 1.2; and (iii) when computing the torque modulus via

\begin{equation*}
    \left|{\boldsymbol\tau}_{{\rm DF}, bg} \right|= |\mathbf{r}|\ \left| \frac{{\rm d}\mathbf{v}_{bg}}{{\rm d}t}\Big\rvert_{\rm DF} \right|,
\end{equation*}

\noindent instead of taking the cross product. In all cases, we obtain a similar evolution for the DF-induced torque and we recover the results described in the next sections.

Figure~\ref{fig:DF_components} shows the time evolution of the quantities used in Equations~\eqref{eq:df_gas}--\eqref{eq:df_stars} to estimate DF; the bottom panels show the DF-induced torque due to each background component. Gas  gives the dominant contribution for most of the time (84, 55, 73, 79, 89, and 50 per cent of the time for MBHs 1--6, respectively), followed by stars, whereas the dark matter contribution is almost negligible. The only exception is MBH~6, for which dark matter dominates DF 20 per cent of the simulation time: this is due to the fact that the MBH gets ejected in the outskirt of the system. 

The significance of DF is very sensitive to the relative velocity between the MBH and the surrounding medium ($\mathbf{v}_{bg}$). If, for simplicity, we focus on the gas component only, we can appreciate that $|\mathbf{v}_{\rm g}|$ is in the range that allows for both $\mathcal{M}>1$ and $|\mathbf{v}_{\rm g}|< 100 $ km s$^{-1}$ for a large fraction of time (at least before the bar starts to have a significant dynamical impact, at $t\approx840$ Myr). The relatively weak DF in the very first  Myr of the evolution is due to our set-up: the initial relative velocity between the MBH and the gas is close to zero by construction, hence the DF torque  magnitude is low, as the motion is initially subsonic.

\subsubsection{Dynamical friction and global torques: a direct comparison}\label{sec:directcompDFtorque}

\begin{figure}
\centering
\includegraphics[ width=.45\textwidth]{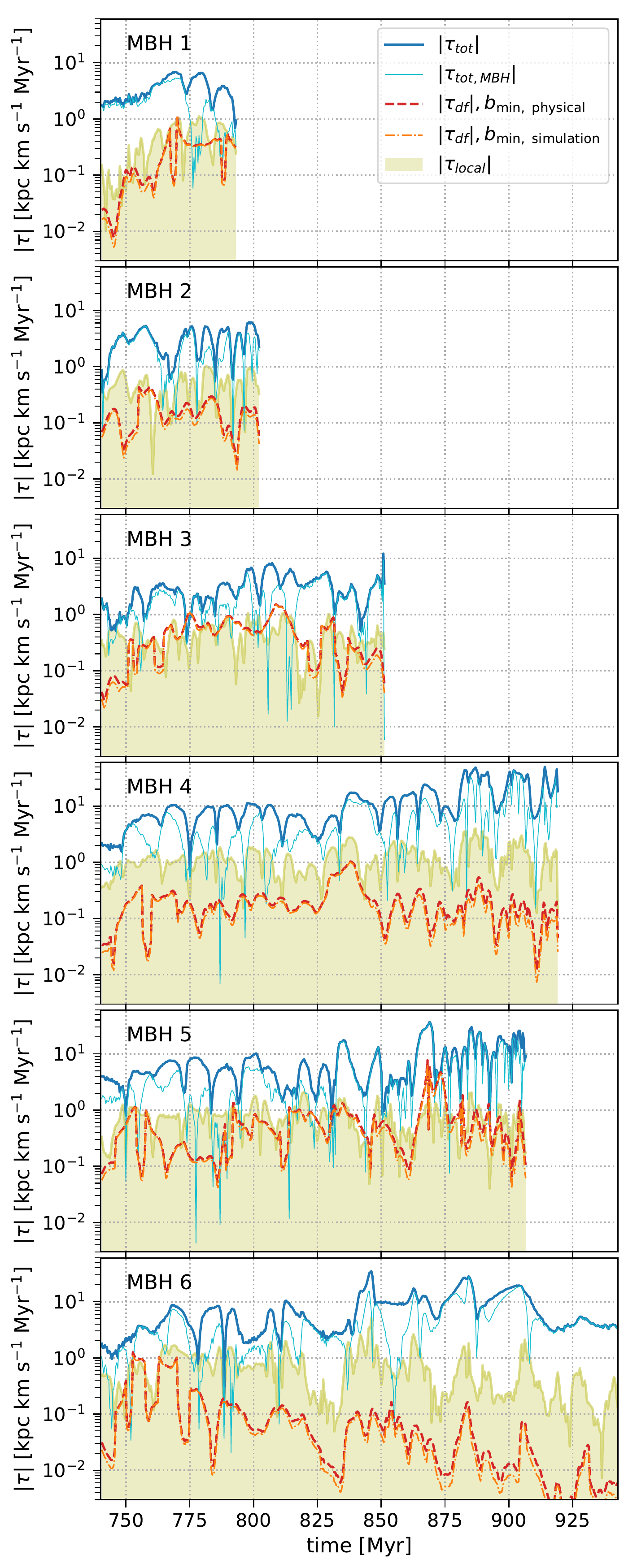}
    \caption{Time evolution of the total gravitational torque modulus (solid, dark blue thick line) and its component along the MBH angular momentum (solid, light blue thin line); the torque is computed directly from the particles in each snapshot. The plot also shows the modulus of the torque associated to the DF, obtained by summing the contribution of the three background components. DF is computed using either the physically motivated value for the minimum impact parameter (Equation~\ref{eq:bmin}; dashed red line) or by assuming an impact parameter equal to the minimum distance resolved in the simulation (dash-dotted orange line; see the text for more details). The upper limit of the yellow-shaded region shows the modulus of the  local torque, as defined in the text. }
    \label{fig:Torque_vs_DF}
\end{figure}

Figure~\ref{fig:Torque_vs_DF} compares the global gravitational torque experienced by each MBH to the DF torque obtained by the sum of the three contributing background species; in what follows, we will always refer to the DF-induced torque as the sum of these  components. Beware that the global torque is computed from all particles, including the closest ones, thus it intrinsically includes the DF acting in the simulation.

It is important to note that the DF might be under-resolved in the simulation owing to its limited resolution. To verify this, we compare the DF computed either adopting the physical minimum impact parameter $b_{\rm min}$ given in Equation~\eqref{eq:bmin} or the maximum between the physical $b_{\rm min}$  and minimum separation that can be resolved in the simulation; the physical $b_{\rm min}$  ranges from a few to $\sim$20~pc for gas, and it is of the order of 1~pc for stars and dark matter;  the minimum resolved separation in the simulation amounts to nearly 10 pc.\footnote{In fact, this value is computed by performing a mass-weighted average of the softening of the interacting particle and the one of the MBH (10 pc), as this is what happens in the integrator.} 
We find that we under-resolve DF by $\approx 10$--$20$ per cent only, 
as shown in Figure~\ref{fig:Torque_vs_DF}. In the remaining of the paper, we will always refer to DF computed adopting the physical $b_{\rm min}$ (Equation~\ref{eq:bmin}).

Figure~\ref{fig:Torque_vs_DF} clearly shows that the global torque experienced by each MBH is typically much larger than that from the DF; on average, the ratio between the global and DF torque amounts to 10--100 for MBHs~1--5 and to $\sim$300 for MBH~6. This is true in the case we consider the \textit{magnitude} of the global gravitational torque (dark blue), and also when we compute the torque \textit{component along the MBH angular momentum direction} (cyan); the latter happens to be very small for very short time ranges, and it is intrinsically more noisy, as it oscillates between positive and negative values.
This fact suggests that the DF is not the main culprit driving the orbital evolution of the MBHs, which seems to be triggered by the inhomogeneities in the large-scale matter distribution, such as the bar and the spiral structures that develop in the galaxy (see Figure~\ref{fig:snapshots}). 

Such consideration is further supported by the analysis of the local torque ${\boldsymbol\tau}_{\rm local}$. We define this quantity as the torque originating from a sphere around each MBH with twice the radius enclosing $2M_2$. 
We find that the magnitude of the local torque (also shown in Figure~\ref{fig:Torque_vs_DF}) is on average a factor of 10--20 smaller than the global one, meaning that the torque acting on each MBH is not dominated by the MBH DF wake. The local torque is sometimes notably larger than the DF torque; to explain this, note that ${\boldsymbol\tau}_{\rm local}$ is estimated within a sphere of typical radius 100--150~pc; however, such radius can become $\approx 400$~pc as the MBH wanders in a low-density region (note in particular MBH~6); when this happens, the sphere's radius is close to or larger than the disc scale-height, and the large density difference within the sphere  may dominate the torque estimate. Also note that we are showing the modulus of the local torque and not its component along the MBH angular momentum.

\begin{figure}
\centering
\includegraphics[width=.45\textwidth]{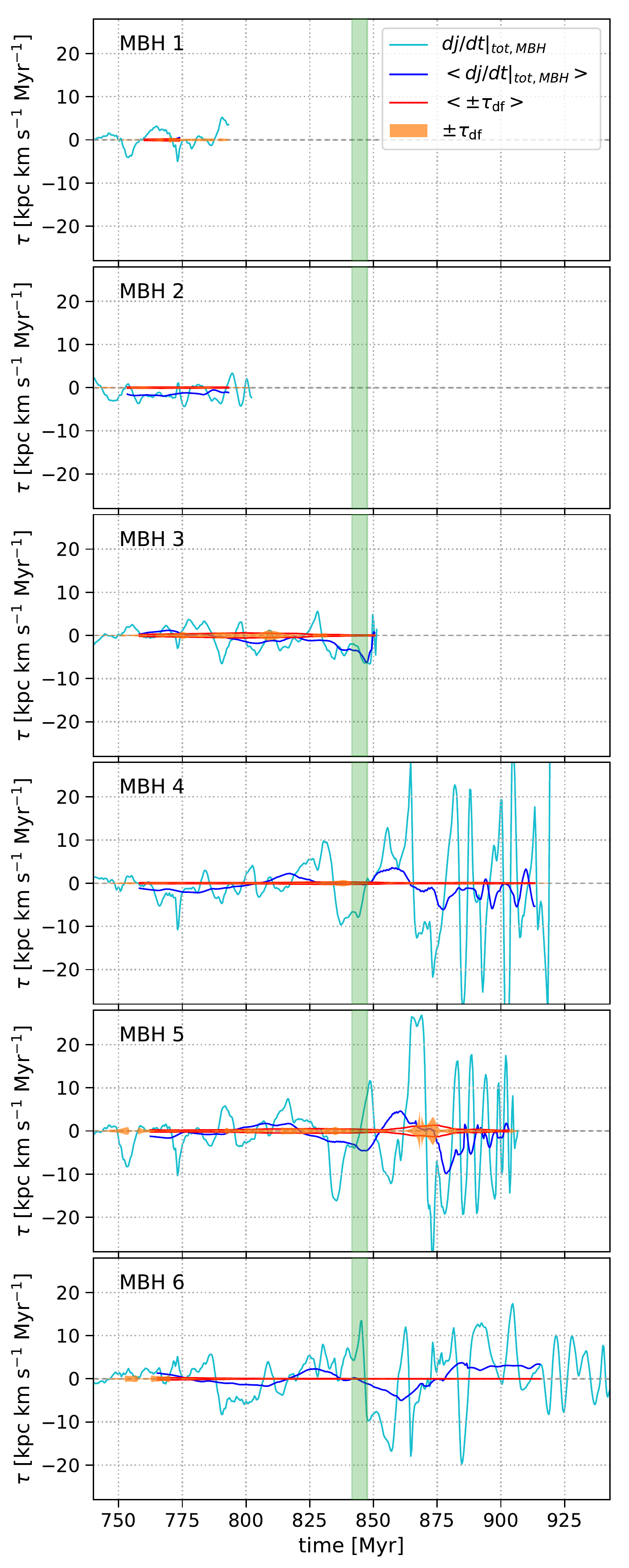}
    \caption{Time evolution of the total gravitational torque projected along the MBH angular momentum direction (cyan  line); here the torque is obtained as the time derivative of the MBH angular momentum, ${\rm d}\mathbf{j}/{\rm d}t$. The blue line shows the moving average of the latter quantity, averaged over a full azimuthal oscillation. The orange region shows the modulus of the DF-induced torque ${\boldsymbol\tau}_{\rm DF}$, obtained as the sum of the three contributing species: the coloured area spans from $-|{\boldsymbol\tau}_{\rm DF}|$ to $|{\boldsymbol\tau}_{\rm DF}|$; the moving average of the same quantity averaged over a full azimuthal oscillation is shown in red. The green vertical line marks the bar formation epoch, as detailed in Section\ref{sec:bar}.}
    \label{fig:Lin_Torque}
\end{figure}

A legitimate question is whether the effect of global torques averages out over a full orbit. To investigate this aspect, we computed the torque as the time derivative of each MBH's angular momentum.%
\footnote{We checked that the measure of the global torque obtained via these two strategies is similar, even if not exactly matching.}  
This latter approach would not be justified if we were to directly compare which effect is more relevant at each instant of time (as in Figure~\ref{fig:Torque_vs_DF}, as the derivative of the angular momentum implicitly assumes we are integrating the global torque over the frequency of our snapshots). However, computing the torque as  the time derivative of the angular momentum is a better strategy  when one wants to average the torque over a certain time-scale.  Figure~\ref{fig:Lin_Torque} shows the time evolution of the  global gravitational torque component along the MBH angular momentum and the associated moving average (computed over a full azimuthal oscillation); the plot also shows the DF-induced torque, both instantaneous and averaged over a full azimuthal oscillation. 
While the global gravitational torque averages to a value close to zero prior to the bar formation, the same is not true after $t\approx 840$~Myr; this is the time at which a stable barred structure starts to appear, and coincides with an enhancement in the torque experienced by each MBH. The figure suggests that the final orbital evolution of MBHs 4--6, and most probably even 3, has to be attributed to the presence of the bar, that produces a significant torque and determines the final fate of the MBHs. 

\subsection{Bar analysis and bar-induced torque}\label{sec:bar}

\begin{figure*}
\centering
\includegraphics[ width=0.95\textwidth]{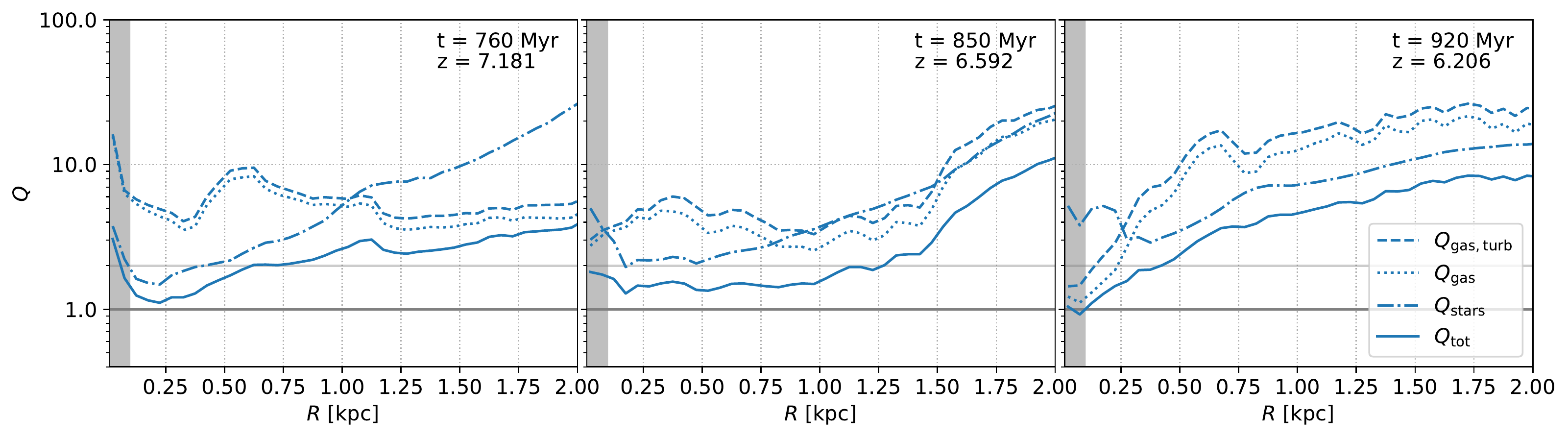}
    \caption{Toomre parameter $Q$ evaluated at three different snapshots of the simulation for stars (dash-dotted line), gas assuming $V=c_{\rm s}$ (dotted line), gas assuming $V=\sqrt{c_{\rm s}^2+\sigma_{\rm r,g}^2}$ (turbulent gas; dashed line) and $Q_{\rm tot}$ (solid line), evaluated via Equation~\eqref{eq:Qtot}. More details on the computation of $Q$ can be found in the text. The horizontal grey lines mark Toomre parameters equal to  1 and 2. The shaded vertical grey area marks the innermost 0.1~kpc, where the effect of softening cannot be neglected.}
    \label{fig:toomre}
\end{figure*}

In order to better understand the development of the bar that dominates the dynamical evolution of the MBH (Figure~\ref{fig:Lin_Torque}), we characterize the stability of the disc in terms of the Toomre parameter,

\begin{equation}
    Q = \frac{\kappa V}{w G \Sigma},
\end{equation}

\noindent where $\Sigma$ is the gas or stellar surface density, $\kappa = \sqrt{3 (V_{\rm t}/R)^2}$ is the epicyclic frequency defined via the azimuthal velocity $V_{\rm t}$ and the cylindric radius $R$ (this approach is analogous to what has been done in e.g. \citealt{Fiacconi2017, Ceverino2010}); 
$w$ is a constant equal to $\pi$ for gas and to 3.36 for stars; finally, $V$  is equal to the radial velocity dispersion $\sigma_{r, \rm stars}$ for stars, to the speed of sound $c_{\rm s}$ for gas and, if the gas is turbulent, it is better approximated by $V=\sqrt{c_{\rm s}^2+\sigma_{\rm r,g}^2}$ where $\sigma_{\rm r,g}$ is the radial velocity dispersion of gas. We also compute the total Toomre parameter, evaluated as
\begin{align} \label{eq:Qtot}
    Q_{\rm tot}^{-1} &= \left\{
\begin{array}{lc}
u Q_{\rm stars}^{-1} + Q_{\rm gas}^{-1} & Q_{\rm stars} \geq Q_{\rm gas} \\
Q_{\rm stars}^{-1} + u Q_{\rm gas}^{-1} & Q_{\rm stars} < Q_{\rm gas}\end{array} \right\} \nonumber;\\ 
    u &= 2 \frac{c_{\rm s} \ \sigma_{r, \rm stars}}{ c_{\rm s}^2 + \sigma_{r, \rm stars}^2 },
\end{align}
where  $Q_{\rm stars}$ ($Q_{\rm gas}$) is the stellar (non-turbulent gaseous) Toomre parameter  \citep{Romeo2013, Fiacconi2017}. 
The Toomre parameter as a function of radius is shown in Figure~\ref{fig:toomre} for three different snapshots. $Q$ is computed in a disc of radius 2~kpc and half-height 0.8~kpc; we did not consider stars belonging to the small nucleus around the primary MBH in the computation, as this component is dispersion dominated.\footnote{Stars belonging to the nucleus have been selected as the stars closer than 0.3~kpc from the primary MBH whose  circularity $|j_z/j_{\rm c}|<0.5$; here $j_z$ is the angular momentum in the $z$ direction and $j_{\rm c}$ the angular momentum of a circular orbit with the same radius. The angular momenta and radii here are computed in a reference frame centred on the position and velocity of the primary MBH.}

The Toomre parameter $Q$ associated to gas is large, suggesting that gas is not in a gravito-turbulent state, as already broadly detailed in \citet{Fiacconi2017}. The total and stellar Toomre parameters, instead, are only slightly larger than one: this hints to a nearly bar-unstable disc that can likely be triggered toward the formation of a large scale structure. In fact, a necessary, but not sufficient, condition for the formation of a bar is that $Q < 2$--3 (e.g. \citealt{BinneyTremaine2008}); whether the bar forms or not depends on other structural parameters of the galactic disc and halo, as well as on numerical details such as the force and time resolutions  \citep[e.g.][]{Debattista2006,Dubinski2009a}.

The bar that develops in our simulation is clearly visible both in the stellar and in the gaseous projected density profiles. This suggests that the triggering for the bar could have been external, as gaseous bars are not likely to form in absence of an external perturbation \citep{Mayer2004}. However, a much deeper analysis would be required to address the genesis of the bar: this aspect will be explored in a forthcoming paper.

\begin{figure}
\centering
\includegraphics[ width=.4\textwidth]{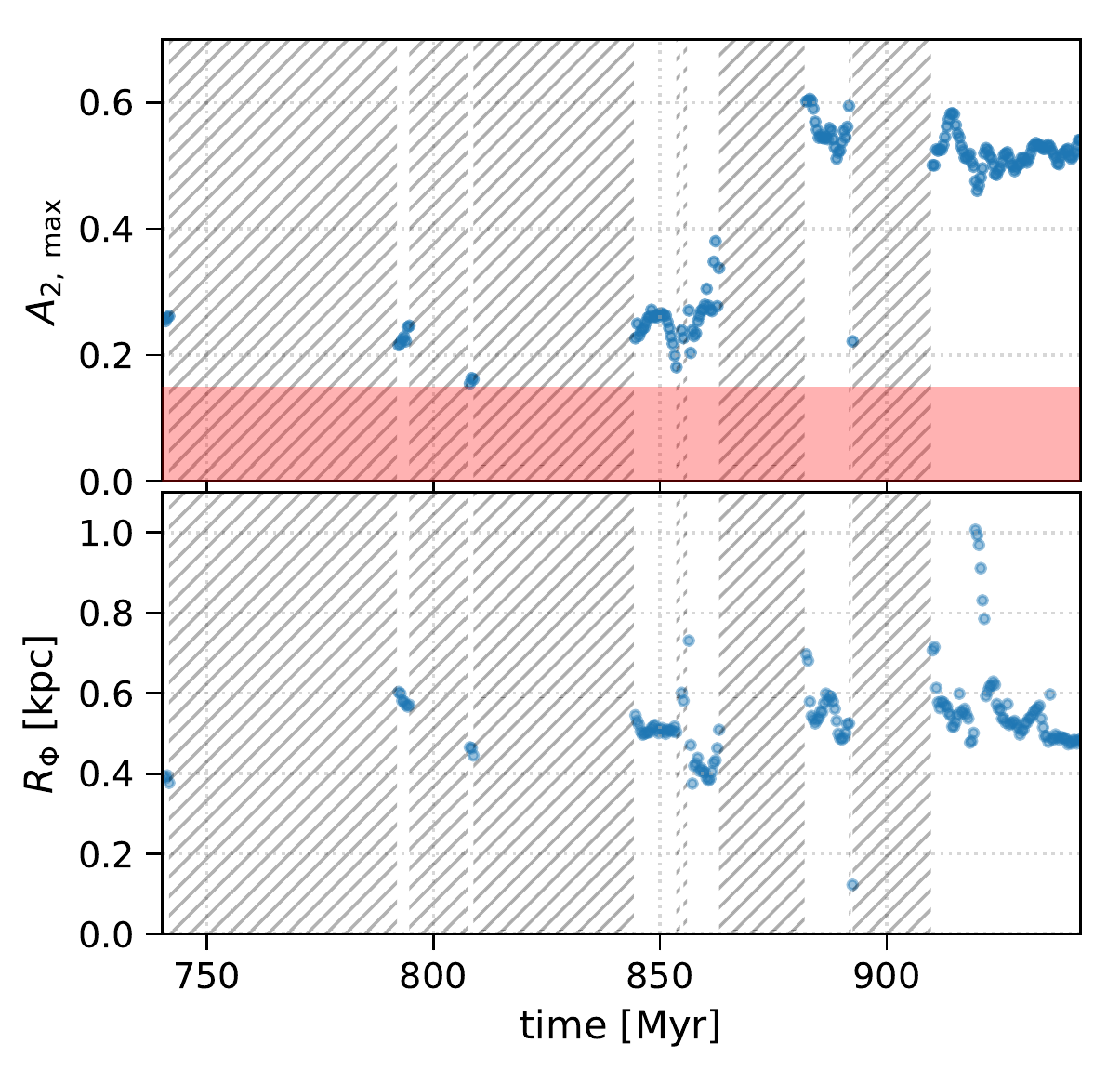}
    \caption{Time evolution of the bar properties. Upper panel: bar strength $A_{2, \rm{max}}(R)$, bottom panel: bar length $R_{\Phi}$ (see text for details). The horizontal, red stripe in the upper panel marks the threshold $A_{2, \rm{max}}(R) = 0.15$.
    The grey shaded areas mark the time intervals in which we are unable to retrieve the bar parameters because of our conservative constraints.}
    \label{fig:bar}
\end{figure}

In order to give a quantitative representation of the bar structure that is growing in the system, we perform a Fourier decomposition of the stellar surface density distribution $\Sigma_{*}$, following in detail the process described in \citet{Zana19}.
In particular, we compute the ratio between the second and the zeroth components of the Fourier development
\begin{equation}
    A_2(R) \equiv \frac{\left|\sum_{j}{m_{j}e^{2i\theta_{j}}}\right|}{\sum_{j}m_{j}},
    \label{eq:A2}
\end{equation}
and the angular phase of the overdensity, as
\begin{equation}
    \Phi(R) \equiv \frac{1}{2} \arctan \left[ \frac{\sum_{j}{m_{j}\sin(2\theta_{j})}}{\sum_{j}{m_{j}\cos(2\theta_{j})}} \right],
    \label{eq:phase}
\end{equation}
where the summation is carried over all the $j$-th stellar particles within an annulus of width 2~pc, height 1~kpc and centred at the cylindrical radius $R$. $\theta_{j}$ is the azimuthal angle of the $j$-th particle of mass $m_j$.

When the stellar disc hosts a barred overdensity over a certain radial range, the  parameter $A_{2}(R)$ reaches a peak near the outer edge of the structure and the value of $A_{2, \rm max} \equiv {\rm max}[A_{2}(R)]$ provides an estimate for the bar strength.
Another outcome of the Fourier decomposition which is significant for our analysis is the parameter $A_2(<R)$, evaluated through Equation~\eqref{eq:A2}, but including all the particles that are enclosed within the radius $R$.
In this study we compute the length of the bar by checking the fluctuation of the phase $\Phi(R)$.
If $R_{\rm peak}$ is the radius where the function $A_2(<R)$ has its local maximum, we define the bar extent, $R_{\Phi}$, as the radius $R$ of the last annulus where the condition \mbox{$\left| \Phi(R_{\rm peak})-\Phi(R)\right|<\arcsin(0.15)$} is valid, thus the phase of the stellar overdensity starts to vary significantly.

The time evolution of the parameters $A_{2, {\rm max}}$, and $R_{\Phi}$ is shown in Figure~\ref{fig:bar}.
Given the chaotic nature of the high-redshift disc, with developing substructures and disturbances from the external environment, 
the assessment of the bar parameters is not straightforward.
In order to avoid the noise due to the unsteady environment, we do not take into account any deviation from axisymmetry with $A_{2, {\rm max}}(R)<0.15$.
For this reason both the strength and length of the bar show significant ranges where it is impossible to retrieve their values unless a much deeper analysis is carried out, which is beyond the purpose of this work. 

The bar seed could be already growing around $t=750$~Myr  but it is only after about $840$~Myr  that a stable structure is in place [$A_{2, {\rm max}}(R)>0.15$] -- this epoch is marked in Figure~\ref{fig:Lin_Torque} in order to relate the raise of the torque magnitude with the presence of the bar. The bar rapidly reaches a strength $A_{2 \ \rm max} \simeq 0.5$ and a length of about $R_{\Phi}\simeq0.5$~kpc, and maintains these values till the end of the run.

\section{Discussion}\label{sec:discussion}

\subsection{Stochasticity generators}

Our analysis so far clearly suggests that the stochastic orbital evolution of the MBHs is to be attributed to the global torques arising from the whole host system, rather than DF alone. In the previous sections, we highlighted that the presence of the bar significantly contributes to the global torques. The torques that cannot be attributed to the bar itself (or the torques acting prior to the bar formation) have to be associated to some other perturbations in the potential. 

We checked where the global gravitational torques mainly come from, at each snapshot of the simulation: most of them originate from the bar and spiral structures, whereas the impact of the central overdensity in which the primary MBH is embedded (which oscillates around the galactic centre for most of the evolution) seems not to have a major impact, unless the separation between the inspiralling and the primary MBH is closer than $\sim$0.2~kpc. We note that galactic fly-bys may also give a relevant contribution to the overall torque.

We should also mention that, to some level,  DF and global torques are two sides of the same coin. Stochasticity is at least partially due to the fact that the MBHs are not evolving in a smooth background with relatively fixed properties; instead, the turbulent status of the medium implies that, at each time, the MBH finds itself in a background with a different sound speed, density, temperature; as a consequence, the significance of the DF itself is intrinsically stochastic. In short, gravitational inhomogeneities are sources of stochasticity as they induce random net torques, but the same inhomogeneities also randomise the significance of DF.

A legitimate question is to which degree the significant stochastic torques acting in the presently discussed high-redshift galaxy are in place at later epochs, and especially in the present day Universe. On the one hand, it is clear that high-redshift systems are more chaotic and violent environments compared to their present-day counterparts \citep[e.g.][]{Mortlock2013}, thus the overall significance of their stochastic torques is probably much larger than today. On the other hand, in this study we see most of the torque regulating the MBHs orbital evolution coming from the bar and associated spirals. Such structures are abundant in the local Universe (\citealt{Kelvin2018} -- a prominent example is our own Milky Way, \citealt{Portail2017}) and presumably at all times.
It is thus reasonable to expect that the stochastic orbital evolution of MBHs, driven by bars and spiral structures, occurs at all cosmic epochs.

A further relevant point is the fact that the stochasticity in the MBHs orbital evolution occurs in a galaxy that is not featuring massive star-forming clumps, which could also induce an erratic evolution \citep{Tamburello2017, Roskar2015}. Galaxies featuring massive clumps appear to be very common in the Universe at $z\approx 1$--3 \citep{Shibuya2016}; however, their occurrence is much rarer both in the local Universe and at the redshifts we are exploring here  \citep{Shibuya2016}. The lack of massive clumps in our system is possibly related to the fact that the large specific star formation rate ($\sim 5 $ Gyr$^{-1}$) in main-sequence galaxies at $z>5$, and the associated stellar feedback, maintains the gas in a turbulent status; in addition, the relatively low galaxy mass prevents the formation of large gravitational instabilities in the gaseous component, thus the formation of massive clumps (see \citealt{Fiacconi2017} for more details on this aspect).

\subsection{Orbital evolution and DF time-scale}

\begin{table}
  \centering
  \caption{Comparison between the inspiral time-scale inferred from DF and the effective inspiral time in the simulation.}
  \label{tab:timescale}
  \begin{center}
  
\begin{tabular}{cccccc}
\hline
MBH & $j_0$ & $T_0$ & $\tau_{{\rm DF},0}$ & $j_0/\tau_{{\rm DF},0}$ & $t_{\rm inspiral}$ \\ 
\#  & [$\frac{\rm kpc\ km}{\rm s}$] & [Myr] & [$\frac{\rm kpc\ km}{\rm s\ Myr}$]  & [Myr]  &  [Myr] \\ \hline
1 & 87  &  39   & 0.13  & 653 & >60 \\
2 & 87  &  34   & 0.11  & 757 & 50  \\
3 & 87  &  35   & 0.21  & 415 & 110 \\
4 & 160 &  45   & 0.11  &1419 & 140 \\
5 & 160 &  44   & 0.22  & 718 & 160 \\
6 & 160 &  52   & 0.23  & 710 & $\gg$210 \\ \hline
\end{tabular}
 
  \end{center}
\justifying
{\footnotesize For each MBH (first column), the table shows the magnitude of the initial angular momentum $j_0$ (second column), the time $T_0$ to complete the first full azimuthal oscillation   (third column), the average DF torque over $T_0$ (fourth column), the time-scale associated to the DF (fifth column), and the effective time for the inspiral (sixth column). The reported time-scales suggest that the DF is not the main responsible for the orbital decay.}
\end{table}

The time-scale for the MBH orbital decay is the key parameter that encodes the efficiency of the MBH pairing. Table~\ref{tab:timescale} compares the effective decay time of each MBH to the decay computed as $j_0/\tau_{\rm DF,0}$, where $j_0$ is the modulus of the MBHs initial angular momentum and $\tau_{\rm DF,0}$ is the magnitude of the torque associated to the DF, averaged over the first full azimuthal oscillation $T_0$ (also in the table). The inspiral time one would infer relying on the DF only is a factor 4--15 larger than the actual inspiral time for the MBHs that manage to reach the centre of the system.\footnote{Note that a more precise estimate for the DF-driven decay time would account for the fact that the DF-torque likely increases while the angular momentum decreases and the orbit shrinks. Because of this, the actual DF inspiral time could be shorter, but we expect our approach to provide a good estimate, as most of the decay time is expected to be spent at large radii.} 
The final angular momentum of MBH~6 is twice its initial, and the DF it experiences by the end of the run is $\sim$100 times smaller than the initial one: its inspiral time is thus most likely going to exceed $j_0/\tau_{\rm DF,0}$  by a large amount, unless global torques play in its favour in the subsequent evolution. In this respect, it is worth mentioning that a galaxy merger with mass ratio $\sim$1:5 is about to happen ($\approx 50$ Myr after the end of our integration the intruder enters the disc of the main galaxy) and the bar gets severely trimmed when this happens (see also \citealt{Guedes2013}), thus it may not influence the further orbital evolution of MBH~6.

\subsection{Bars in the young Universe} 
What is clear from our analysis 
is that, once the bar forms, it heavily affects the orbital evolution and it is the most prominent torque source, together with the spirals. 
Given this, it would be important to infer the fraction of strong bars in the early Universe that LISA is going to probe. The fraction of barred galaxies amounts to $>30$ per cent in the present day Universe \citep[][and references therein]{Consolandi16}. %
Although it is much harder to obtain similar statistics for the high-redshift Universe, numerical experiments suggest that the fraction of galaxies harbouring a bar is significant ($\approx40$ per cent; \citealt{Algorry2017, Rosas-Guevara2020}) at $z\sim 1$, thus it is reasonable to expect that such fraction would remain significant at larger redshift ($z \sim 7$--6).

In general, high-redshift galaxies are expected to suffer much more frequent dynamical interactions due to external perturbations compared to low-redshift ones.
Although external perturbations seem not to be critical for the formation of a stellar bar, as highlighted by recent cosmological zoom-in simulations, some specific combinations of impact parameter and relative velocity between the main galaxy and an external perturber could boost the bar formation in potentials that are already bar-unstable (see \citealt{Zana18a} for a recent discussion).
Note that high-redshift bars could be frequently killed or trimmed by the frequent merger events and fly-by passages, therefore resulting short-lived. Nevertheless, they could re-appear multiple times if the external perturbations do not deeply modify the galactic potential, as found in \citet{Zana18b}.
Therefore, concerning our case, the reason for the sudden weakening 
of the structure for a few Myr around $t = 875$ and 900 Myr  (see Figure~\ref{fig:bar}) could be due to some internal or external perturbations to the disc, that are strong enough to 
undermine the structural integrity of the bar, but 
not enough to ultimately alter the potential well.

\subsection{Small-scale decay via stellar hardening}

\begin{figure}
\centering
\includegraphics[ width=.35\textwidth]{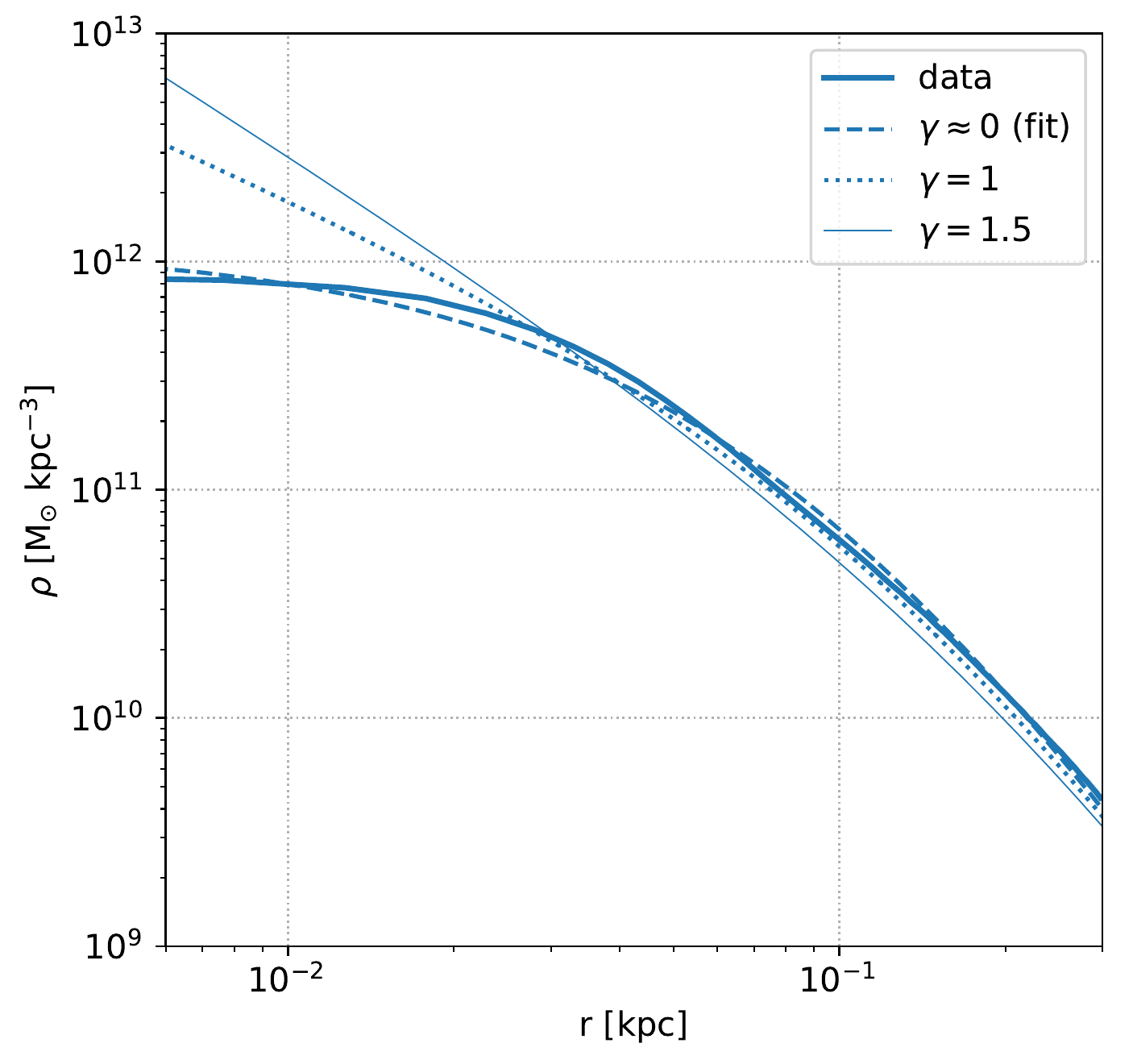}
    \caption{Spherical density profile of the stars surrounding the primary MBH near the end of our run, at $t=920$ Myr. The thick solid line shows the density profile as obtained from the run, the dashed line shows the best fit obtained using a \citet{Dehnen1993} density profile, while the dotted and thin solid lines show the extrapolation of the same profile in the assumption of a inner cusp $\rho\propto r^{-\gamma}$ with respectively $\gamma = 1, 1.5$. Details on the fit and on the extrapolated profiles can be found in the text. }
    \label{fig:density_profile}
\end{figure}

After the MBHs reach a pericentre of less than 100~pc, their dynamics cannot be resolved in our simulation. However, the fact that the primary MBH is surrounded by a dense nucleus dominated by stars allows us to  infer an upper limit for the decay time down to the GW emission phase.

The  MBHs need to get close enough to form a hard 
MBH binary, i.e. a binary whose semimajor axis is smaller than $a_{\rm h} = GM_2/(4\sigma^2)$ \citep[][]{Merritt2013} which, in our case is $\sim$0.1~pc if $\sigma$ is computed from stars in a sphere of 150~pc around the primary. Stars dominate the density profile at any time within 150~pc from the primary  (the stellar mass fraction here is 60--80 per cent). Therefore, it is reasonable to consider stars as the sole contributors to the DF (the presence of a gaseous circumnuclear disc would probably boost the inspiral, \citealt{Mayer2007}, although see \citealt{Souza-Lima2017,Souza-Lima2020}). In the nucleus, the stellar motion is isotropic to a good approximation, so that the average relative speed between the MBH and the medium coincides with the velocity dispersion. We can thus express the time-scale for the last segment of the DF-driven inspiral to be

\begin{equation}
    t_{\rm DF\star} \approx 76 \  \frac{7 }{\ln \Lambda}  \left( \frac{r_{\rm e}}{0.1\ \rm kpc}\right)^2 \left( \frac{\sigma}{100 \rm\ km\, s^{-1}}\right) \left( \frac{10^6 {\rm M}_{\sun{}}}{M_2}\right){\rm Myr};
\end{equation}

\noindent here $r_{\rm e}$ is both the scale radius of the central stellar nucleus (whose properties are discussed below) and the separation at which the small-scale DF inspiral starts; $r_{\rm e}$ here is also used as the maximum impact parameter entering in the Coulomb logarithm, while the minimum one is set to 0.1 pc. The velocity dispersion around the primary MBH increases from about 50 to 150~km~s$^{-1}$ throughout the simulation, meaning that the DF-driven decay is of order $\sim$100~Myr. 
Afterwards, the binary separation could be reduced by three-body scatterings. The time-scale to reach the final coalescence can be estimated following \citet{Sesana2015}:

\begin{align} \label{eq:thard}
    t_{\rm gw} &\approx \frac{\sigma_{\rm infl}}{GH\rho_{\rm infl}a_{\rm */gw}},\nonumber\\ 
    a_{\rm */gw}^5 &= \frac{64G^2M_1M_2(M_1+M_2)\sigma_{\rm infl}}{5c^5H\rho_{\rm infl}}f(e),
\end{align}

\noindent where $\rho_{\rm infl}$ and $\sigma_{\rm infl}$ are the stellar density and velocity dispersion, respectively, within the binary influence radius $r_{\rm infl}$ (i.e. the sphere containing twice the binary mass in stars), $H\approx 15$ is a parameter obtained via scattering experiments \citep{Quinlan1996,Sesana2006}, $c$ is the speed of light in vacuum, and $f(e)=(1-e^2)^{-7/2}(1+73/24e^2+37/96e^4)$ is the eccentricity ($e$) enhancement function for GW emission \citep{Peters1964}. The length-scale $a_{\rm */gw}$ is the typical separation at which the binary shrinking  transits from being dominated by three-body scatterings to being mainly  driven by GWs and practically represents the separation at which a MBH binary spends most part of the time during the stellar hardening evolution.

\begin{table}
  \centering
  \caption{Time-scales for stellar hardening and GW-induced coalescence}
  \label{tab:decay_time}
  \begin{center}
  
\begin{tabular}{lcccccc}
\hline
 &  $t_{\rm inspiral}$ & $t_{\rm inspiral}$ & $t_{\rm inspiral}$ \\
 $\gamma$ & $e=0$  & $e=0.5$ &  $e=0.9$ \\ \hline
 $\approx0$   &  585 Myr &  426 Myr &  141 Myr \\
 $1.0$        &  220 Myr &  160 Myr &  53 Myr \\
 $1.5$        &   87 Myr &   63 Myr &   21 Myr \\\hline
\end{tabular}

  \end{center}
\justifying
{\footnotesize The table shows the typical time-scale needed to reach the final coalescence once a hard MBH binary is formed. The time-scales are obtained via Equation~\eqref{eq:thard}. The first column shows the inner density slope $\gamma$ of the assumed stellar density profile; the second, third, and fourth columns show the decay time-scale for MBH binaries whose eccentricity in the transition between stellar hardening and GW-decay is respectively equal to 0, 0.5, and 0.9. Details on the assumed profiles can be found in the text and in Figure~\ref{fig:density_profile}.}
\end{table}

Figure~\ref{fig:density_profile} shows the density profile of stars around the primary MBH near the end of the integration ($t=920$~Myr). We fit it using a \citet{Dehnen1993} density profile,

\begin{equation}
    \rho(r)=\frac{(3-\gamma)M_{\rm t}}{4\pi}\frac{r_0}{r^\gamma(r+r_0)^{4-\gamma}},
\end{equation}

\noindent and obtain a total mass $M_{\rm t}=4.3\times10^9$~M$_{\sun{}}$, a scale radius $r_0 =0.1$~kpc, and an inner density slope $\gamma \approx 0$; the fitted profile is also shown in Figure~\ref{fig:density_profile}. If one estimates the velocity dispersion via $\sigma_{\rm infl}^2=G(M_1+M_2)/[(1+\gamma)r_{\rm infl}]$ \citep{Alexander2005}, the decay time-scale obtained via Equation~\eqref{eq:thard} would amount to the values shown in the first line of Table~\ref{tab:decay_time}. We can note that {\it only} a large binary eccentricity\footnote{Note that in \citet{Sesana2015}, the eccentricity is assumed to remain constant for the whole evolution, generally predicting a faster decay, especially when GWs become dominant. Nevertheless, this could severely affect the final result only when the initial eccentricity is quite large, i.e. $1-e \ll 0.1$.} at the onset of the GW phase guarantees a decay time shorter than the previous evolutionary time-scales. 

In fact, the limited resolution of the simulation does not allow us to properly resolve the density profile in the innermost 100~pc, which is likely steeper than what we found. For this, we extrapolated the inner density slope by using a \citet{Dehnen1993} density profile with $M_{\rm t}$ obtained from the previous fit, but imposing an inner slope $\gamma=1$ and 1.5, and ensuring in both cases that the radius containing 0.1$M_{\rm t}$ is the same in the fitted and extrapolated profiles. The extrapolated profiles are shown in Figure~\ref{fig:density_profile}. The decay time-scales obtained via  Equation~\eqref{eq:thard} for the extrapolated profiles are also listed in  Table~\ref{tab:decay_time}. The decay time-scale is in these cases of the order of or shorter than the decay time of the previous stages, meaning that the coalescence would supposedly happen promptly once the hard binary is formed. Nonetheless, the hardening and inspiral time-scales we inferred in Table~\ref{tab:decay_time} are definitely longer than the $\approx 15$~Myr found by \citet{Khan2016}, who addressed the pairing of $\sim 10^8$~M$_{\sun{}}$ MBHs in a massive early type galaxy at $z\approx 3$, which would become a BCG by $z\approx 0$.
The discrepancy in the time-scales has to be attributed to the large central density in their merging systems, which is nearly $10^5$~M$_{\sun{}}{\rm pc}^{-3}$ at 10 pc (their figure 6), i.e. $\sim$100 times larger than in our case \citep[see also the discussion in][]{Khan2018}.\footnote{In fact, Equation~\eqref{eq:thard} results in a time-scale of 5--25~Myr if we adopt a stellar density 100 times larger than that in Figure~\ref{fig:density_profile}, the MBH masses in \citet{Khan2016} ($M_1=3\times 10^8$~M$_{\sun{}}$, $M_2=8\times 10^7$~M$_{\sun{}}$), and we rescale $\sigma_{\rm infl}$ to these MBH masses.}
Note that the system they explored is not a main-sequence galaxy as the one presented in this paper, but instead it is a more massive and relatively rare galaxy. For this reason, we expect the time-scales we obtain in this paper to be more representative for the bulk of the galaxy population.

Note that the hardening and inspiral time-scales extrapolated for our system do not account for the effect of gas (which is expected to assist the orbital decay at scales larger than a pc, but could hinder it at smaller scales; \citealt{Souza-Lima2020}), and the possibility that the fragmentation of a non resolved circumnuclear disc could render even this smaller-scale decay very stochastic \citep{Fiacconi2013,Souza-Lima2017}. 
If we neglect these complications, our estimates suggest that the whole pairing process could take as short as a few hundred Myr and probably less than a Gyr for the majority of the studied MBHs; that is, the MBH merger is likely going to occur at $z = 5$--3.5, except for the MBH that  undergoes the bar-induced ejection. Concerning the fate of this latter object, one could wonder how likely this scattering event is. Even if our statistical sample is too limited to make strong statements here, it is worth mentioning that a test simulation with virtually the same initialization as the current one also showed evidence for  a bar slingshot, suggesting that the event can be relatively common. 

\subsection{Caveats and limitations}

Our simulation has the important limitation that MBH accretion and feedback were not taken into account. However, it is reasonable to expect that feedback would render DF even less efficient by evacuating the gas mass around an MBH (see, e.g. \citealt{Park_Bogdanovic_2017,Souza-Lima2017,delValle2018} for a smaller scale analysis of this aspect). Each inspiralling MBH may also enhance its mass by accreting gas along the decay: in this respect, it is worth remembering that the MBH masses we adopted in the current run are relatively large for $z\gtrsim 6$, thus we do not expect this to represent a major issue. We also note that accretion on to MBHs may be used in future high-redshift surveys to sample offset or dual active galactic nuclei \citep[e.g.][]{Capelo_et_al_2017,DeRosa_et_al_2019}, in order to estimate how many MBHs could have been ejected by non-axisymmetric structures such as the bar described in this work.

A further simplistic assumption we made is that MBHs are initially placed on circular orbits aligned with the disc plane. This choice was made in order to perform a relatively controlled numerical experiment, in which we can better disentangle  different processes. As a matter of fact, the typical galaxy orbit at the onset of a galaxy merger is the parabolic one \citep[e.g.][]{Benson2005}; in fact, orbits are expected to get more circular, prograde \citep[e.g.][]{Bonetti2020} and in-plane \citep[e.g.][]{Walker1996, Laine1999} along the decay in a disc-like, rotationally-supported system as ours, so that our initial configuration could be eventually reached. In addition, it is difficult to predict whether inclined orbits would be more or less affected by stochastic torques, as both DF and global torques are likely less effective out of the disc plane. 
Furthermore, it is important to consider that stochastic torques randomize the MBH orbits relatively quickly in the present simulation, rendering them eccentric and at least mildly off-plane. Thus, although our initial conditions are idealized, the MBHs configuration gets quickly altered, resembling a more realistic physical situation.

Another important caveat to mention is that, in spite of the unprecedented spatial and force resolution we adopt in this zoom-in simulation, we might still slightly underestimate DF. In order to better dissect this aspect, let us make some instructive   comparison with \citet{Tremmel2015}. They show that an external term for DF needs to be introduced in the integrator in order to properly model this process, and they compare simulations with and without the semi-analytical treatment for DF. Their most resolved run features a force softening and an MBH mass close to ours, but a typical particle mass more than $10$ times larger than the gas particle mass in our run: with this set-up, they find that the DF-induced decay gets shorter by a small factor ($\approx$1.5) if they switch on the semi-analytical DF prescription. Their result implies that we likely do not significantly underestimate DF in our (better resolved) run. Even neglecting these considerations, the secondary MBH masses we adopted here are deliberately large compared to what one would expect for a minor merger at $z\approx 7-6$, implying that the DF experienced by a typical intruder MBH at that redshift would be weaker; if one wants to be more conservative, our secondary MBHs could be thought as having an `effective' mass of a few$\times10^5\msun$. 

The magnitude of DF along our run could even be over-estimated owing again to resolution limits that do not allow to properly resolve gaseous structures at sub-pc scales (see e.g. \citealt{Souza-Lima2020}); however, in this case the importance of large scale torques would be even more prominent.

Finally, the semi-analytical approach we develop to estimate the significance of DF is intrinsically approximate, and Eq.~\ref{eq:df_stars} could perform poorly in some situations. In particular, in collisionless environments, the adopted  DF treatment is not adequate in galaxies featuring extended cores \citep{Read2006, Petts2016}, and more in general in systems in which fast moving stars are crucial for a proper DF estimate. However, the main source of DF in the present framework is gas, which does not suffer analogous limitations and whose effect is more local \citep{Ostriker1999}. 

\section{Summary and conclusions}\label{sec:concl}

In this paper, we studied the orbital evolution of a set of MBHs inspiralling in a typical star-forming, turbulent non-clumpy galaxy embedded in a cosmological box at $z = 7$--6. The primary and secondaries have masses of, respectively, $4\times 10^6$ and $10^6$~M$_{\sun{}}$, implying that their coalescence would shine in the LISA GW detection band with extremely large signal-to-noise. The secondaries are assumed to be brought in the main galaxy via a minor merger that completely strips the intruder galaxy. Here we summarize our key results:

\begin{enumerate}

    \item The evolution of the secondary MBH's orbit is highly stochastic and the time for the inspiral can vary significantly even for MBHs starting from the same initial distance to the centre, but with a different orbital phase;
    
    \item the dynamical-friction (DF) induced torque appears to be weaker at least by a factor of a few (but generally by orders of magnitude) compared to the global gravitational torque experienced by each inspiralling MBH, implying that the DF should not be regarded as the main driver for the orbital decay, at least in the high-redshift Universe;
    
    \item the formation of a bar (and associated spiral structure) in the main galaxy crucially influences the orbital decay: it appears to boost the inspiral for 2--3 secondary MBHs, whereas in one case it ejects an MBH in the galaxy outskirts. Thus, bars and spiral arms are crucial sources of stochasticity for the orbital decay: they overall appear to boost the inspiral, even though in some cases they may ultimately hinder it;
    
    \item the effective decay time for 4 out of 5 inspiralling MBHs is a factor 4--15 shorter than what would be inferred relying on the DF alone;
    
    \item the whole pairing process, including the final shrinking due to stellar interactions, occurs on time-scales not much shorter than the age of the Universe at $z \approx 6$, and the final coalescence  is likely going to occur at $z = 5$--3.5 (except for the case in which the bar ejects the inspiralling MBH).
    
\end{enumerate}

The possibility of an erratic orbital decay is not completely new to the study of MBH inspiral: as already mentioned, giant star-forming clumps have been proposed as further sources of stochasticity \citep{Fiacconi2013,Roskar2015, Tamburello2017, Souza-Lima2017}; clumpy galaxies have been suggested to be typical environments at $z\approx 1$--3, but their occurrence at the epoch addressed here is likely much lower \citep{Shibuya2016}. In addition, it is unclear whether the blobs found in observations of galaxies at $z \sim 1$--2, that have been long interpreted as giant clumps, are instead under-resolved smaller object \citep[e.g.][]{Ivison2020}.
The existence of bars and spiral structures is obvious even in the low-redshift Universe, thus we believe the source of stochasticity in the current study to be more general, and potentially acting at all redshifts.

To conclude, our results clearly indicate that the standard (\citealt{Chandrasekhar1943} and \citealt{Ostriker1999}) prescriptions that model  DF as a local drag force in analytical and semi-analytical models can be greatly inadequate once the cosmological environment is taken into account, and especially in the young Universe that LISA is going to probe. A large effort is thus needed in this direction, in order to infer realistic time-scales for the inspiral in the presence of bars, spiral arms, and perturbations from external galaxies, as such disturbances may ultimately determine the final fate of an inspiralling MBH. 

\section*{Acknowledgements}
We warmly thank Douglas Potter for making available a tool for the computation of accelerations via {\textsc{pkdgrav3}}. We thank the referee for their useful comments and suggestions, and Michael Tremmel, Fazeel Khan, Priyamvada Natarajan and Lumen Boco for fruitful discussion.
EB, PRC, and LM acknowledge support from the Swiss National Science Foundation under the Grant 200020\_178949.

\section*{Data Availability Statement}
The data underlying this article will be shared on reasonable request to the corresponding author.


\bibliography{bibliography} 

\bsp	
\label{lastpage}
\end{document}